# Generalized coherent states in thermal field dynamics in the frame of diagonal ordering operator technique

## Dušan POPOV[a, b]

[a] University Politehnica Timisoara, Department of Physical Foundations of Engineering, B-dul Vasile Pârvan No. 2, 300223 Timisoara, Romania

[b] Serbian Academy of Nonlinear Sciences (SANS), Kneza Mihaila 36, Beograd-Stari Grad, Belgrade, Serbia

E-mail: dusan_popov@yahoo.co.uk   dusan.popov@upt.ro

ORCID: https://orcid.org/0000-0003-3631-3247

**Abstract**

Although the thermofield dynamics (TFD) was developed specifically for systems to which the ladder canonical operators $\hat{a}^+$ and $\hat{a}$ are associated, in the paper we showed that this approach can also be formulated for systems of deformed bosons, associated with a pair of deformed operators $\hat{\mathcal{A}}_+$ and $\hat{\mathcal{A}}_-$. In this context we built sets of temperature dependent coherent states, both of the Barut-Girardello and of the Klauder-Perelomov type, these two types being dual to each other. In calculations involving ladder operators, both independent and temperature-dependent, we used the Diagonal Operator Ordering Technique (DOOT), which is a generalization of the Integration Within an Ordered Product technique (IWOP) applicable to canonical operators associated with the one-dimensional harmonic oscillator. We also refer to the connection between TFD and thermodynamic quantities.

## 1. Introduction

The idea of thermofield dynamics (TFD) appeared for the first time in the 8th decade of the last century, with the aim of explaining some temporal aspects related to non-equilibrium systems. It was due to Umezawa and collaborators, who later extended the idea to collective thermal phenomena of condensed states [1], [2]. Later, the idea attracted the attention of other researchers who formulated the theory of thermal phenomena embracing the formalism of coherent states [3]. In essence, TFD is a formalism in quantum field theory and in fields like high-energy physics, quantum optics, and condensed matter physics that extends the theory formulated from zero temperature phenomena to corresponding phenomena involving finite temperatures. Later, the TFD used quantum phase distributions [4], as well as Man and Revzen used TFD to define the thermal coherent states [3]. The thermal states representation were suitable for introducing some thermal entities, e.g. the thermal Wigner operator, useful in quantum optics [5]. The idea was revived a few years later when specific notions of TFD were used to obtain Husimi's thermo operator, widely used in squeezed states in quantum optics [6]. This enriched the use of the TFD method and also opened a new path in solving master equations. Using the basic ideea and notions of TFD, this method was also formulated for time-dependent systems, both for bosons and for fermions [7]. Also, in the frame of TFD were investigated some non-dissipative phenomena at certain finite low temperatures involving atoms in the cavity mode which are initially in the thermal equilibrium state and thermal CSs [8]. Thermofield approach can also be successfully applied in quantum chemistry [9]. A variant of generalized coherent states for the harmonic oscillator system has also been formulated using the time-dependent TFD. This is based on Glauber's formulation of coherent states, where the authors introduce two types of time-dependent coherent states: coherent thermal (CT) and thermal coherent (TC), which are related to each other [10].

All these listed applications, as well as others, in various fields of physics show that the TFD approach is considered one of the most appropriate for the study of phenomena in which the quantities involved depend on temperature or time.

On the other hand, the foundations of the coherent states (CSs) formalism were laid as early as 1926 through Schrodinger's paper [11], with the intention of finding solutions of the Schrödinger equation that satisfy the correspondence principle. After a period of reduced interest from researchers in this problem, a real informational boom occurred around the middle of the last century. During that period, papers and books on CSs appeared. Among them, we mention only the fundamental ones [12], [13], [14], [15], [16] and the references therein. These led to the formulation of three different ways of defining CSs: 1.) Barut-Girardello way, in which CSs are eigenvectors of the annihilation operator [17]; 2.) Klauder-Perelomov way, in which CSs are defined as the result of the action of the quantum displacement operator on a ground state (vacuum) [18] and 3.) Gazeau-Klauder way, in which CSs are parametrized by classical canonical action–angle variables [19].

For the linear quantum harmonic oscillator all three definitions are convergent leading to the same mathematical expression for CSs, based on the result of the action of the canonical generation $\hat{a}^+$ and annihilation $\hat{a}$ operators. For this reason, CSs associated with the linear quantum oscillator are also called linear CSs or canonical CSs. However, since many physical phenomena have a nonlinear character, it has proven necessary to construct appropriate CSs, called nonlinear CSs. For these kinds of CSs, so-called nonlinear or deformed generalized creation $\hat{\mathcal{A}}_+$ and annihilation $\hat{\mathcal{A}}_-$ operators were used, which involved a real function that depends on the particle number operator $\hat{n}$, called the deformation function $f(\hat{n})$. The deformation function is not unique, but has different expressions for the various quantum systems studied. There are many papers related to nonlinear CSs, some of the most relevant being [20], [21], [22], [23]. Therefore, the synergistic approach between the two approaches, TFD and CSs, is interesting, which is also one of the goals of this paper.

In this paper we have limited ourselves to examining the TFD and CSs only to describe systems of deformed bosons.

## 2. Preliminaries about the TFD

The thermofield dynamics (TFD) model or formalism is a theoretical framework for studying quantum field theories at finite temperatures. In this formalism description of thermal effects and phenomena at temperature $T \to 0$ is extended to respective phenomena at non-zero temperatures $T \neq 0$, $T = (k_B \beta)^{-1}$, where $k_B$ is the Boltzmann constant, and $\beta = (k_B T)^{-1}$ is the temperature parameter. This extension is done in such a way as to preserve the structure of the operator algebra, as well as the operator's commutation relations. In non-zero temperatures quantum mechanics, the expectation value or thermal average of a "normal" operator $\hat{O}$ is expressed as an ensemble average, using the normalized density operator $\hat{\rho}(\beta)$ attached to the mixed states

$$<\hat{O}>_T = \mathrm{Tr}\left[\hat{O}\hat{\rho}(\beta)\right] = \frac{1}{Z(\beta)}\sum_{n=0}^{\infty}<n|\hat{O}e^{-\beta\hat{\mathcal{H}}}|n> = \frac{1}{Z(\beta)}\sum_{n=0}^{\infty}e^{-\beta E_n}<n|\hat{O}|n> \quad (2.1)$$

On the other hand, in the TFD theory, the physical quantities at equilibrium are represented by their thermal expectation values. The central idea in TFD is to express this ensemble average as an expectation value over *just one state* $|\Psi(\beta)> \equiv |0(\beta)>$, called the thermofield double vacuum state or the thermal vacuum state.

For any Hermitian operator $\hat{O}_H$ acting in the physical Hilbert space $\mathsf{H}$, the thermal vacuum is defined as [3]. In this context, the earlier expectation value can be written as

$$<\hat{O}_H>_T = \frac{<0(\beta)|\hat{O}_H|0(\beta)>}{<0(\beta)|0(\beta)>} = \frac{\mathrm{Tr}\left(\hat{O}_H e^{-\beta\hat{\mathcal{H}}}\right)}{\mathrm{Tr}\left(e^{-\beta\hat{\mathcal{H}}}\right)} \quad (2.2)$$

The primary entity from which the TFD formalism starts is just this temperature-dependent state, which we called in the next the thermal vacuum state $|0(\beta)> \in \mathsf{H}$. This state belongs to an original or physical Hilbert space $\mathsf{H}$. Takahashi and Umezava defined the thermal vacuum $|0(\beta)>$ at finite temperature $T \neq 0$, such that the expected value of an operator $\hat{O}_H$ with respect to the thermal equilibrium will be equivalent to the statistical mean value with respect to the thermal state [1]:

$$<0(\beta)|\hat{O}_H|0(\beta)> \equiv <\hat{O}_H>_T \quad (2.3)$$

Although this condition seems to be of the orthogonality type, it is not satisfied by c-numbers. Consequently, the thermal vacuum state $|0(\beta)>$ cannot be an element of the original Hilbert space [24]. This was the reason why Takahashi and Umezawa resorted to the old idea to enlarge the Hilbert space, i.e. to introduce an additional Hilbert space – called the "tilde space" which is the complex conjugate copy of the original zero temperature Hilbert space [1].

Since in thermofield dynamics (TFD), in addition to dynamic degrees of freedom, thermal degrees of freedom are also examined, then, in addition to ordinary particles, so-called tilde particles are also introduced. In other words, the ordinary particles possess the dynamical degrees of freedom, while the fictitious (tilda) particles possess the thermal degrees of freedom. The former represent the dynamic degrees of freedom, and those in the second category the thermal ones. For this purpose Y. Takahashi and H. Umezawa introduced an additional, "doubled" or fictitious Hilbert space (or called a tilde-conjugate field) denoted $\tilde{H}$, coupled with the physical space through a direct product, so that the whole or extending Hilbert space is a tensor product $H_{tot} = H \otimes \tilde{H}$ [1]. The whole Hilbert space is spaned by the direct product of the eigenstates of physical Hamiltonian $\hat{\mathcal{H}}$ and those of the new introduced tilde Hamiltonian $\hat{\tilde{\mathcal{H}}}$ which has the same energy eigenvalues as their counterpart $\hat{\mathcal{H}}$, i.e. $\hat{\tilde{\mathcal{H}}} |\tilde{n}> = E_{\tilde{n}} |\tilde{n}>$.

So, the original optical field state $|n>$ in the space $H$ is accompanied by a tilde state $|\tilde{n}>$ in $\tilde{H}$. The same correspondence is also applicable to operators: any operator $\hat{O}_H$ acting on $H$ has a corresponding or image $\hat{\tilde{O}}_{\tilde{H}}$ in $\tilde{H}$. If the $H$ and $\tilde{H}$ are two separable Hilbert spaces associated with the subsystems $S$ (of the real particles), respectively $\tilde{S}$ (of the so called tilde particles), which are uncorrelated, then the composite Hilbert space $H_{total}$, associated with the whole system $S_{total}$ is given by the tensor product, $H_{total} = H \otimes \tilde{H}$. The total Hamiltonian of the whole (or composite) system $S_{total}$ (both physical and fictitious systems) is also the tensor product of two subsystems $\hat{\mathcal{H}}_{total} = \hat{\mathcal{H}} \otimes \hat{\tilde{\mathcal{H}}} = \hat{\mathcal{H}} \otimes \hat{I} + \hat{I} \otimes \hat{\tilde{\mathcal{H}}}$. Alternatively, it is not difficult to prove the validity of the following equality: $\exp(-\beta \hat{\mathcal{H}}_{total}) = \exp(-\beta \hat{\mathcal{H}}_H) \otimes \exp(-\beta \hat{\tilde{\mathcal{H}}}_{\tilde{H}})$.

By extending the usual Hilbert space with a "doubled" space, a thermal state (which is a mixed state) can be represented as a pure state in the total space. Therefore, by calculating an expected value of a pure state created by TFD, we can obtain a thermal average in usual statistical mechanics. Since a mixed state is a statistical ensemble of pure states at finite temperature, by using the TFD method, the statistical average of an observable at finite temperature will actually be the expectation value of the pure state.

According to the quantum Liouville equation, a composite and isolated system with whole Hilbert space $H \otimes \tilde{H}$ is always in the pure state with the corresponding density operator $\hat{\rho}_n \equiv |n, \tilde{n}><n, \tilde{n}| = |n><n| \otimes |\tilde{n}><\tilde{n}|$. This always commute with the Hamiltonian of the physical system $\hat{\mathcal{H}} = \sum_m E_m |m><m|$.

$$\left[\hat{\mathcal{H}}, \hat{\rho}_n\right] |s> = \hat{\mathcal{H}} \hat{\rho}_n |s> - \hat{\rho}_n \hat{\mathcal{H}} |s> =$$
$$= \sum_m E_m |m><m|n><n|s> - |n><n|\sum_m E_m |m><m|s> = E_s |s> - E_s |s> = 0 \quad (2.4)$$

i.e. this result agree with the quantum Liouville equation. Hence, the whole vector basis in the composite system must be $|n> \otimes |\tilde{n}> \equiv |n, \tilde{n}> \in H \otimes \tilde{H}$, and the double vacuum (or ground) state for both physical and fictitious modes, can be written as $|0, \tilde{0}>$.

As a partial conclusion, let us emphasize that the thermal vacuum state $|0(\beta)>$ is a pure (only!) state of the composite system (physical and tilde). Then, according to Schmidt decomposition and purification, there is an orthonormal basis $\{|n>\}$ for $\mathsf{H}$ and $\{|\tilde{n}>\}$ for $\tilde{\mathsf{H}}$, such that

$$|0(\beta)>\equiv \sum_n C_n |n>\otimes|\tilde{n}>=\sum_n C_n |n,\tilde{n}> \qquad (2.5)$$

where the coefficients $C_n$, called Schmidt coefficients, are real non-negative numbers, satisfying the condition $\sum_n C_n^2 = 1$. This procedure, named purification, allows us to associate pure states with mixed states.

Consequently, for each Hilbert space it exists the following density operators, with identical eigenvalues

$$\hat{\rho}_{\mathsf{H}} = \sum_n C_n^2 |n><n| \quad , \quad \hat{\tilde{\rho}}_{\tilde{\mathsf{H}}} = \sum_n C_n^2 |\tilde{n}><\tilde{n}| \qquad (2.6)$$

Finally, for a single-mode, the connection within the vacuum thermal state (which play the role of ground state in the whole Hilbert space) and the double energy basis, in TFD the Schmidt coefficients are defined as $C_n = \sqrt{\dfrac{e^{-\beta E_n}}{Z(\beta)}}$ [2] and the thermal vacuum is

$$|0(\beta)>= \sqrt{\dfrac{e^{-\beta \hat{\mathcal{H}}}}{Z(\beta)}} \sum_n |n,\tilde{n}> = \sum_n \sqrt{\dfrac{e^{-\beta E_n}}{Z(\beta)}} |n,\tilde{n}> \qquad (2.7)$$

where the partition function

$$Z(\beta) = \mathrm{Tr}\, e^{-\beta \hat{\mathcal{H}}} = \sum_n e^{-\beta E_n} \qquad (2.8)$$

ensures normalization to unity of the states.

The complex conjugate expression for the thermal vacuum state is

$$<0(\beta)|= \sum_n <n,\tilde{n}|\sqrt{\dfrac{e^{-\beta \hat{\mathcal{H}}}}{Z(\beta)}} = \sum_n \sqrt{\dfrac{e^{-\beta E_n}}{Z(\beta)}} <n|\otimes<\tilde{n}| \qquad (2.9)$$

The projector of the thermal vacuum double state is, therefore

$$\hat{\rho}_{|0(\beta)>} \equiv |0(\beta)><0(\beta)|= \sum_n \sqrt{\dfrac{e^{-\beta E_n}}{Z(\beta)}} |n,\tilde{n}> \otimes \sum_m <m,\tilde{m}|\sqrt{\dfrac{e^{-\beta E_m}}{Z(\beta)}} \qquad (2.10)$$

The density operator corresponding to the dynamical degree of freedom, i. e. in the physical Hilbert space, is obtained by calculating the partial trace over the whole thermal vacuum state, i.e. over the thermal vacuum projector $\hat{\rho}_{|0(\beta)>} \equiv |0(\beta)><0(\beta)|$ [10]:

$$\mathrm{Tr}_{\tilde{\mathsf{H}}}\, \hat{\rho}_{|0(\beta)>} = \mathrm{Tr}_{\tilde{\mathsf{H}}} |0(\beta)><0(\beta)|= \sum_m <\tilde{m}|0(\beta)><0(\beta)|\tilde{m}>=$$

$$= \sum_m <\tilde{m}\left[\sum_n \sqrt{\dfrac{e^{-\beta E_n}}{Z_H(\beta)}}|n>\otimes|\tilde{n}> \cdot \sum_j \sqrt{\dfrac{e^{-\beta E_j}}{Z_H(\beta)}}<j|\otimes<\tilde{j}|\right]|\tilde{m}>= \qquad (2.11)$$

$$= \sum_{n,j} \sqrt{\dfrac{e^{-\beta E_n}}{Z_H(\beta)}}\sqrt{\dfrac{e^{-\beta E_j}}{Z_H(\beta)}}|n>\otimes<j|\cdot \delta_{nj} = \dfrac{1}{Z_H(\beta)}\sum_n e^{-\beta E_n}|n><n|=\hat{\rho}_{\mathsf{H}}(\beta)$$

As will be seen below, this is just the expression of the thermal density operator that corresponds to the thermal equilibrium state of a quantum system, whose states are not pure but mixed. This result is in agreement with the fact that a quantum mechanical system which is in the state of thermal equilibrium at finite temperature $T$ cannot be found in a pure state, but it is in a mixed state described by a density matrix [25]. The normalization to unity of the thermal vacuum lead to the expression of partition function

$$<0(\beta)|0(\beta)> = \sum_{n,m} \sqrt{\frac{e^{-\beta E_n}}{Z(\beta)}} \sqrt{\frac{e^{-\beta E_m}}{Z(\beta)}} (<m|\otimes<\tilde{m}|)(|n>\otimes|\tilde{n}>) = \frac{1}{Z(\beta)} \sum_{n=0}^{\infty} e^{-\beta E_n} = 1 \quad (2.12)$$

In the thermofield dynamics (TFD) formalism, the thermal vacuum $|0(\beta)>$, at finite temperature is defined in such a manner that the vacuum expectation value of an operator, which act on the physical Hilbert space $\mathsf{H}$, $<0(\beta)|\hat{O}_\mathsf{H}|0(\beta)>$, to be equal with their corresponding statistical (or thermal) average:

$$<0(\beta)|\hat{O}_\mathsf{H}|0(\beta)> = <m,\tilde{m}|\sum_m \sqrt{\frac{e^{-\beta E_m}}{Z(\beta)}} \hat{O}_\mathsf{H} \sum_n \sqrt{\frac{e^{-\beta E_n}}{Z(\beta)}} |n,\tilde{n}> =$$

$$= \sum_m \sqrt{\frac{e^{-\beta E_m}}{Z(\beta)}} \sum_n \sqrt{\frac{e^{-\beta E_n}}{Z(\beta)}} <m|\hat{O}_\mathsf{H}|n> \otimes <\tilde{m}|\tilde{n}> = \sum_n \frac{e^{-\beta E_n}}{Z(\beta)} <n|\hat{O}_\mathsf{H}|n> = <\hat{O}_\mathsf{H}>_T \quad (2.13)$$

where $\hat{\rho}(\beta) = \frac{1}{Z(\beta)} e^{-\beta \hat{\mathcal{H}}}$ is the density operator which characterize the mixed state of a canonical system at non-zero temperature, and their properties will be examined below.

So, the above relation shows that the average value of a pure state in a thermal vacuum is identical to the average thermal value of the system in a mixed state at a non-zero temperature. Therefore, a series of methods and results obtained for treating mixed states of systems at non-zero temperatures can also be implemented in the case of thermal vacuum, that is, in TFD.

In TFD, the temperature dependence of various quantities is achived through a temperature dependent parameter $\theta(T) = \theta(\beta)$.

For a free (non-interacting) bosons system, the temperature-dependent parameter $\theta(\beta)$ is defined such that the vacuum expectation value agrees with the statistical average [26].

$$\cosh \theta(\beta) = (1 - e^{-\beta \hbar \omega})^{-\frac{1}{2}} \quad , \quad \sinh \theta(\beta) = (e^{\beta \hbar \omega} - 1)^{-\frac{1}{2}} \quad (2.14)$$

or equivalently

$$\tanh \theta(\beta) = \exp\left(-\beta \frac{\hbar \omega}{2}\right) \quad , \quad \theta(\beta) = \frac{1}{2} \ln \frac{1 + e^{-\beta \frac{\hbar \omega}{2}}}{1 - e^{-\beta \frac{\hbar \omega}{2}}} \quad , \quad \lim_{\substack{T \to 0 \\ (\beta \to \infty)}} \theta(\beta) = 0 \quad (2.15)$$

This obtained expression is characteristic for a boson system associated with the one dimensional harmonic oscillator, because the quantum field for a harmonic oscillator is equivalent to a system of non-interacting bosons (e.g photons), where each oscillation mode correspond to a single boson. Since the energy of a one-dimensional harmonic oscillator is $E_n = \hbar \omega \left(n + \frac{1}{2}\right)$, partition function is $Z(\beta) = e^{-\beta \frac{\hbar \omega}{2}} (1 - e^{-\beta \hbar \omega})^{-1}$ and the expression of the thermal vacuum becomes

$$|0(\beta)>= \sum_{n=0}^{\infty} \sqrt{\frac{e^{-\beta E_n}}{Z(\beta)}} |n, \tilde{n}> = \sum_{n=0}^{\infty} \sqrt{\frac{e^{-\beta \hbar \omega \left(n+\frac{1}{2}\right)}}{Z(\beta)}} |n> \otimes |\tilde{n}> =$$

$$= \sqrt{1 - e^{-\beta \hbar \omega}} \sum_{n=0}^{\infty} \left( e^{-\beta \frac{\hbar \omega}{2}} \right)^n \frac{(\hat{a}^+)^n}{\sqrt{n!}} |0> \otimes \frac{(\hat{\tilde{a}}^+)^n}{\sqrt{n!}} |\tilde{0}> = \sqrt{1 - e^{-\beta \hbar \omega}} \sum_{n=0}^{\infty} \frac{1}{n!} \left( e^{-\beta \frac{\hbar \omega}{2} \hat{a}^+ \hat{\tilde{a}}^+} \right)^n |0, \tilde{0}> =$$

$$= \sqrt{1 - e^{-\beta \hbar \omega}} \exp \left( e^{-\beta \frac{\hbar \omega}{2} \hat{a}^+ \hat{\tilde{a}}^+} \right) |0, \tilde{0}> = \frac{1}{\cosh \theta(\beta)} e^{\tanh \theta(\beta) \hat{a}^+ \hat{\tilde{a}}^+} |0, \tilde{0}>$$

(2.16)

Thus, there is another expression for the thermal vacuum of the system of free bosons

$$|0(\beta)>= \frac{1}{\cosh \theta(\beta)} e^{\tanh \theta(\beta) \hat{a}^+ \hat{\tilde{a}}^+} |0, \tilde{0}> \qquad (2.17)$$

## 3. TFD for deformed bosons

After this general information about the TFD formalism, we are able to extend, as a natural way, the usual definition of coherent states (CSs) to zero-temperature quantum fields and to obtain so called thermal coherent states (TCSs). The algebra of thermal operators is identical to that of operators at zero temperature. This has the consequence that between all the formal properties of the usual CSs and TCSs there is a natural correspondence.

Generalized coherent states (GCSs) are generated by a pair of adjoint (conjugate transpose) creation (raising) $\hat{\mathcal{A}}_+$ and annihilation (lowering) $\hat{\mathcal{A}}_-$ operators, which differ from the corresponding canonical operators $\hat{a}^+$ and $\hat{a}$ (associated with the one-dimensional quantum harmonic oscillator, HO-1D) in that they also contain a nonlinearity function $f(\hat{n})$ that depends on the particle number operator, $\hat{n} = \hat{a}^+ \hat{a}$, where $\hat{n}|n>=n|n>$:

$$\hat{\mathcal{A}}_+ = \sqrt{f(\hat{n})}\, \hat{a}^+ \quad , \qquad \hat{\mathcal{A}}_- = \hat{a}\sqrt{f(\hat{n})} \qquad (3.1)$$

The nonlinearity function $f(\hat{n})$ is not the same for all quantum systems, it has different expressions for different systems. Consequently, the GCSs are also named nonlinear coherent states (NCSs) because of the presence of function $f(\hat{n})$. This satisfy the eigenvalue equation $f(\hat{n})|n>= f(n)|n>$. Consequently, the structure constants which we denote $\rho(n)$ depend on the deformation function $f(\hat{n})$. At the limit $f(\hat{n}) \to 1$, the NCSs turn into canonical CSs.

Let us we consider that the dimensionless Hamiltonian $\hat{\mathcal{H}}$ of the original (physical) system is $\hat{\mathcal{H}} = \hat{\mathcal{A}}_+ \hat{\mathcal{A}}_-$. Their counterpart in the tilde (fictitioous) Hilbert space for the fictitious bosons is $\hat{\tilde{\mathcal{H}}} = \hat{\tilde{\mathcal{A}}}_+ \hat{\tilde{\mathcal{A}}}_-$. Implicitly we consider here $\hbar \omega = 1$. The pair of ladder operators defined above acts according to the following rule: $\hat{\mathcal{A}}_-$ and $\hat{\mathcal{A}}_+$ acts only on the vectors of the physical Hilbert space H and their complex conjugate Hilbert space H*, while $\hat{\tilde{\mathcal{A}}}_+$ and $\hat{\tilde{\mathcal{A}}}_-$ acts only on the vectors of the tilde Hilbert space $\tilde{H}$ and their adjoins $\tilde{H}^*$. The same rule will also apply to thermal operators, i.e. that is, those who depend on the parameter $\beta$. This idea can be illustrated as follows:

$$\{\hat{\mathcal{A}}_-, \hat{\mathcal{A}}_+, \hat{\mathcal{A}}_-(\beta), \hat{\mathcal{A}}_+(\beta), \hat{O}_H\}\begin{cases}\in H, H^* \\ \notin \tilde{H}, \tilde{H}^*\end{cases}, \quad \{\hat{\tilde{\mathcal{A}}}_-, \hat{\tilde{\mathcal{A}}}_+, \hat{\tilde{\mathcal{A}}}_-(\beta), \hat{\tilde{\mathcal{A}}}_+(\beta), \hat{\tilde{O}}_{\tilde{H}}\}\begin{cases}\notin H, H^* \\ \in \tilde{H}, \tilde{H}^*\end{cases} \quad (3.2)$$

This rule will ensure the normal ordering of operators, imposed by the DOOT technique, that is, the creation operators $\hat{\mathcal{A}}_+$ and $\hat{\tilde{\mathcal{A}}}_+$ are placed on the left, and the annihilation operators $\hat{\mathcal{A}}_-$ and $\hat{\tilde{\mathcal{A}}}_-$ are placed on the right. These operators act on double Fock vectors in the following manner:

$$\hat{\mathcal{A}}_-|n,\tilde{n}\rangle = \sqrt{e(n)}|n-1,\tilde{n}\rangle, \quad \langle n,\tilde{n}|\hat{\mathcal{A}}_+ = \sqrt{e(n)}\langle n-1,\tilde{n}|,$$
$$\hat{\mathcal{A}}_+|n,\tilde{n}\rangle = \sqrt{e(n+1)}|n+1,\tilde{n}\rangle, \quad \langle n,\tilde{n}|\hat{\mathcal{A}}_- = \sqrt{e(n+1)}\langle n+1,\tilde{n}| \quad (3.3)$$
$$\langle n,\tilde{n}|\hat{\mathcal{A}}_+\hat{\mathcal{A}}_-|n,\tilde{n}\rangle = e(n), \quad \langle n,\tilde{n}|\hat{\mathcal{A}}_-\hat{\mathcal{A}}_+|n,\tilde{n}\rangle = e(n+1)$$

and similar for the tilde operators.

It can be noted that these relations are not eigenvalue equations. Therefore, the operators $\hat{\mathcal{A}}_-$ and $\hat{\mathcal{A}}_+$ do not have the Fock vectors as eigenvectors. Consequently, $(\hat{\mathcal{A}}_-)^+ \neq \hat{\mathcal{A}}_+$ means that the operators $\hat{\mathcal{A}}_+$ and $\hat{\mathcal{A}}_+$ are not Hermitian.

In the above relations we used the notation $e(n) \equiv n f(n)$ and now let us choose that this has the following general expression (which will be implicitly motivated in the below) [27]:

$$_pe_q(n) = n \, _pf_q(n) \equiv n \frac{\prod_{j=1}^{q}(b_j - 1 + n)}{\prod_{i=1}^{p}(a_i - 1 + n)}, \quad n = 1, 2, 3, \ldots, \quad (3.4)$$

This would mean that the deformation function has the expression

$$_pf_q(\hat{n}) \equiv \frac{\prod_{j=1}^{q}(b_j - 1 + \hat{n})}{\prod_{i=1}^{p}(a_i - 1 + \hat{n})} \quad (3.5)$$

In the expressions above, $p$ and $q$ are natural numbers, and in the next we use the notations $\boldsymbol{a} \equiv \{a_i\}_1^p \equiv a_1, a_2, \ldots, a_p$ and $\boldsymbol{b} \equiv \{b_j\} \equiv b_1, b_2, \ldots, b_q$ for two sets of real numbers, and also for the structure constants

$$_p\rho_q(n) \equiv \prod_{j=1}^{n}e(j) = n! \frac{\prod_{j=1}^{q}(b_j)_n}{\prod_{i=1}^{p}(a_i)_n} = \frac{\prod_{i=1}^{p}\Gamma(a_i)}{\prod_{j=1}^{q}\Gamma(b_j)} \Gamma(n+1) \frac{\prod_{j=1}^{q}\Gamma(b_j+n)}{\prod_{i=1}^{p}\Gamma(a_i+n)}, \quad \Gamma(\boldsymbol{a}/\boldsymbol{b}) = \frac{\prod_{i=1}^{p}\Gamma(a_i)}{\prod_{j=1}^{q}\Gamma(b_j)} \quad (3.6)$$

where $(x)_n = \Gamma(x+n)/\Gamma(x)$ is the Pochhammer symbol.

The result of the repeated action of raising operator on the vacuum state is

$$\left(\hat{\mathcal{A}}_+\right)^n |0> = \sqrt{_p\rho_q(n)} |n> = \sqrt{n! \frac{\prod_{j=1}^{q}(b_j)_n}{\prod_{i=1}^{p}(a_i)_n}} |n> \qquad (3.7)$$

Consequently, for the original optical field $|n>$, as well as for the tilde field $|\tilde{n}>$ and their complex conjugates we can write

$$|n> = \frac{1}{\sqrt{_p\rho_q(n)}} \left(\hat{\mathcal{A}}_+\right)^n |0> \quad , \quad |\tilde{n}> = \frac{1}{\sqrt{_p\rho_q(\tilde{n})}} \left(\hat{\tilde{\mathcal{A}}}_+\right)^n |\tilde{0}>$$

$$<n| = \frac{1}{\sqrt{_p\rho_q(n)}} <0|\left(\hat{\mathcal{A}}_-\right)^n \quad , \quad <\tilde{n}| = \frac{1}{\sqrt{_p\rho_q(\tilde{n})}} <\tilde{0}|\left(\hat{\tilde{\mathcal{A}}}_-\right)^n \qquad (3.8)$$

respectively for the entire Hilbert space

$$|n,\tilde{n}> = \frac{1}{_p\rho_q(n)} \left(\hat{\mathcal{A}}_+ \hat{\tilde{\mathcal{A}}}_+\right)^n |0,\tilde{0}> \quad , \quad <n,\tilde{n}| = <0,\tilde{0}| \frac{1}{_p\rho_q(n)} \left(\hat{\mathcal{A}}_- \hat{\tilde{\mathcal{A}}}_-\right)^n \qquad (3.9)$$

The general expression of the pure thermal vacuum state will then be

$$|0(\beta)> = \sum_n \sqrt{\frac{e^{-\beta E_n}}{Z(\beta)}} |n,\tilde{n}> = \sum_n \sqrt{\frac{e^{-\beta E_n}}{Z(\beta)}} \frac{\left(\hat{\mathcal{A}}_+ \hat{\tilde{\mathcal{A}}}_+\right)^n}{_p\rho_q(n)} |0,\tilde{0}> \qquad (3.10)$$

Their counterpart is

$$<0(\beta)| = \sum_m <m,\tilde{m}| \sqrt{\frac{e^{-\beta E_m}}{Z(\beta)}} = \sum_m \sqrt{\frac{e^{-\beta E_m}}{Z(\beta)}} <0,\tilde{0}| \frac{\left(\hat{\mathcal{A}}_- \hat{\tilde{\mathcal{A}}}_-\right)^m}{_p\rho_q(m)} \qquad (3.11)$$

The expected value of the normal ordered product of $\hat{\mathcal{A}}_+ \hat{\mathcal{A}}_-$ operators in the thermal vacuum state is (see, Eq. (2.13)).

$$<0(\beta)| \hat{\mathcal{A}}_+ \hat{\mathcal{A}}_- |0(\beta)> = \sum_n \sqrt{\frac{e^{-\beta E_n}}{Z(\beta)}} <n,\tilde{n}| \hat{\mathcal{A}}_+ \hat{\mathcal{A}}_- \sum_m \sqrt{\frac{e^{-\beta E_m}}{Z(\beta)}} |m,\tilde{m}> = \frac{1}{Z(\beta)} \sum_n e^{-\beta E_n} e(n)$$
(3.12)

where we have taken into account that the expression for energy eigenvalues is $E_n = \hbar \omega e(n)$.

$$<0(\beta)| \hat{\mathcal{A}}_+ \hat{\mathcal{A}}_- |0(\beta)> = \frac{1}{Z(\beta)} \sum_n e^{-\beta \hbar \omega e(n)} e(n) = -\frac{1}{Z(\beta)} \frac{\partial}{\partial(\beta \hbar \omega)} \sum_n e^{-\beta \hbar \omega e(n)} =$$

$$= -\frac{\partial}{\partial(\beta \hbar \omega)} \ln Z(\beta) \qquad (3.13)$$

The final result is then

$$<0(\beta)| \hat{\mathcal{A}}_+ \hat{\mathcal{A}}_- |0(\beta)> = -\frac{\partial}{\partial(\beta \hbar \omega)} \ln Z(\beta) = -\frac{1}{\hbar \omega} \frac{\partial}{\partial \beta} \ln Z(\beta) = \frac{1}{\hbar \omega} U(\beta) \qquad (3.14)$$

where $U$ is the internal energy. Considering the limits $\lim_{f(\hat{n})\to 1}\hat{\mathcal{A}}_+ = a^+$ and $\lim_{f(\hat{n})\to 1}\hat{\mathcal{A}}_- = a$, the above relations will correspond to a system of usual bosons having the well-known expression of the energy eigenvalues $E_n = \hbar\omega e(n) = \hbar\omega\left(n+\frac{1}{2}\right)$.

$$<0(\beta)|a^+a|0(\beta)> = \frac{1}{Z(\beta)}\sum_{n=0}^{\infty} e^{-\beta\hbar\omega\left(n+\frac{1}{2}\right)} <n,\tilde{n}|a^+a|n,\tilde{n}> =$$
$$= \left(1-e^{-\beta\hbar\omega}\right)\sum_{n=0}^{\infty} e^{-\beta\hbar\omega n} n = -\left(1-e^{-\beta\hbar\omega}\right)\frac{\partial}{\partial(\beta\hbar\omega)}\frac{1}{\left(1-e^{-\beta\hbar\omega}\right)} = \frac{1}{e^{\beta\hbar\omega}-1} \equiv n_T = \sinh^2\theta(\beta)$$
(3.15)

where we used the expression of the Bose-Einstein distribution law, which expresses the expected number of particles in an energy state $|n,\tilde{n}>$ for Bose–Einstein statistics.

$$n_T = \frac{1}{e^{\beta\hbar\omega}-1} \tag{3.16}$$

This result is identical to that obtained in [7].

Let us recall that the expression for the internal energy for a bosonic gas (or gas of one-dimensional harmonic oscillators) is

$$U_{tot} = N_{tot}U = -N_{tot}\frac{\partial}{\partial\beta}\ln Z(\beta) = N_{tot}\frac{\hbar\omega}{2}\coth\beta\frac{\hbar\omega}{2} = N_{tot}\hbar\omega\left(n_T + \frac{1}{2}\right) \tag{3.17}$$

where the last term represents the zero energy.

So, obviously, this result can also be transferred to statistical thermodynamics. Thus, for example, if we consider a gas of deformed bosons, in thermodynamic equilibrium with the environment and subject to canonical distribution, knowing the value of the canonical partition function $Z(\beta)$, we will be able to define a series of thermodynamic quantities.

E.g., let us indicate, as functions of the expected value of the normal ordered product of operators $\hat{\mathcal{A}}_+\hat{\mathcal{A}}_-$ in the thermal vacuum state, the expressions for the internal energy $U$

$$U = -\frac{\partial}{\partial\beta}\ln Z(\beta) = \hbar\omega <0(\beta)|\hat{\mathcal{A}}_+\hat{\mathcal{A}}_-|0(\beta)> \tag{3.18}$$

as well as the differential equation that the free energy $F$ satisfies

$$\beta\frac{\partial F}{\partial\beta} + F = \hbar\omega <0(\beta)|\hat{\mathcal{A}}_+\hat{\mathcal{A}}_-|0(\beta)> \tag{3.19}$$

Particularly, if the system under examination has a *linear energy spectrum* (e.g. one-dimensional harmonic oscillator, or pseudoharmonic oscillator), with energy eigenvalues $E_n = \hbar\omega n + E_0$, the energy levels being equidistant, the above definition for $\theta(\beta)$ is retained. Consequently, the expression of thermal vacuum becomes

$$|0(\beta)> = \sqrt{1-e^{-\beta\hbar\omega}} \sum_n \frac{\left(e^{-\beta\frac{\hbar\omega}{2}}\hat{\mathcal{A}}_+\hat{\tilde{\mathcal{A}}}_+\right)^n}{{}_p\rho_q(n)}|0,\tilde{0}> =$$
$$= \sqrt{1-e^{-\beta\hbar\omega}} \; {}_pF_q\left(a\,;\,b\,;\,e^{-\beta\frac{\hbar\omega}{2}}\hat{\mathcal{A}}_+\hat{\tilde{\mathcal{A}}}_+\right)|0,\tilde{0}> = \frac{1}{\cosh\theta(\beta)}\,{}_pF_q\left(a\,;\,b\,;\,\tanh\theta(\beta)\hat{\mathcal{A}}_+\hat{\tilde{\mathcal{A}}}_+\right)|0,\tilde{0}>$$

(3.20)

Using the newly introduced generalized operators, $\hat{\mathcal{A}}_+$ and $\hat{\tilde{\mathcal{A}}}_+$, the thermal vacuum state can be defined similarly as for the canonical operators [28].

Within the TFD theory, in the extended Hilbert space (physical plus tilde) the thermal vacuum equilibrium state for non-interacting bosons, can be expressed in a compact exponentially squeezed form, by a pure state, related to the zero-temperature standard vacuum $|0, \tilde{0}>$, in the following form

$$|0(\beta)> = \frac{1}{\cosh\theta(\beta)} e^{\tanh\theta(\beta)\hat{\mathcal{A}}_+\hat{\tilde{\mathcal{A}}}_+} |0, \tilde{0}> = \left(1-e^{-\beta\hbar\omega}\right)^{\frac{1}{2}} \exp\left(e^{-\beta\frac{\hbar\omega}{2}}\hat{\mathcal{A}}_+\hat{\tilde{\mathcal{A}}}_+\right)|0, \tilde{0}> \quad (3.21)$$

Their complex conjugate expression is

$$<0(\beta)| = \frac{1}{\cosh\theta(\beta)} <0, \tilde{0}| e^{\tanh\theta(\beta)\hat{\mathcal{A}}_-\hat{\tilde{\mathcal{A}}}_-} = \left(1-e^{-\beta\hbar\omega}\right)^{\frac{1}{2}} <0, \tilde{0}| \exp\left(e^{-\beta\frac{\hbar\omega}{2}}\hat{\mathcal{A}}_-\hat{\tilde{\mathcal{A}}}_-\right) \quad (3.22)$$

Consequently, the thermal vacuum state projector is

$$|0(\beta)><0(\beta)| = \frac{1}{\cosh^2\theta(\beta)} \# e^{\tanh\theta(\beta)\hat{\mathcal{A}}_+\hat{\tilde{\mathcal{A}}}_+} |0, \tilde{0}><0, \tilde{0}| e^{\tanh\theta(\beta)\hat{\mathcal{A}}_-\hat{\tilde{\mathcal{A}}}_-} \# \quad (3.23)$$

According to the DOOT rules (the motivation will be presented below), the projector of the double vacuum $|0, \tilde{0}><0, \tilde{0}|$ is, then

$$|0, \tilde{0}><0, \tilde{0}| = (|0><0|)\left(|\tilde{0}><\tilde{0}|\right) = \frac{1}{\#_pF_q\left(\boldsymbol{a};\boldsymbol{b};\hat{\mathcal{A}}_+\hat{\mathcal{A}}_-\right)\#} \frac{1}{\#_pF_q\left(\boldsymbol{a};\boldsymbol{b};\hat{\tilde{\mathcal{A}}}_+\hat{\tilde{\mathcal{A}}}_-\right)\#} \quad (3.24)$$

The final expression of the thermal vacuum state projector is

$$|0(\beta)><0(\beta)| = \frac{1}{\cosh^2\theta(\beta)} \# \frac{e^{\tanh\theta(\beta)\hat{\mathcal{A}}_+\hat{\tilde{\mathcal{A}}}_+}}{_pF_q\left(\boldsymbol{a};\boldsymbol{b};\hat{\mathcal{A}}_+\hat{\mathcal{A}}_-\right)} \# \# \frac{e^{\tanh\theta(\beta)\hat{\mathcal{A}}_-\hat{\tilde{\mathcal{A}}}_-}}{_pF_q\left(\boldsymbol{a};\boldsymbol{b};\hat{\tilde{\mathcal{A}}}_+\hat{\tilde{\mathcal{A}}}_-\right)} \# \quad (3.25)$$

Following Fan Hong-Yi [5] we can now introduce the thermal vacuum state at ultra-high temperature ($T\to\infty$, $\beta\to 0$)

$$\lim_{\substack{T\to\infty\\\beta\to 0}}|0(\beta)> = \lim_{\substack{T\to\infty\\\beta\to 0}}\sum_n \sqrt{\frac{e^{-\beta E_n}}{Z(\beta)}}\sqrt{\frac{e^{-\beta E_n}}{Z(\beta)}}|n, \tilde{n}> = \sum_n |n, \tilde{n}> \quad (3.26)$$

where $\sum_n |n,\tilde{n}> = \sum_n |n>\otimes|\tilde{n}> \equiv |I>$ play the role of unit operator.

This is, in fact, expression of the thermal pure vacuum state at ultra-high temperature.

Let us note that the vacuum state $|0, \tilde{0}>$ is annihilated by either $\hat{\mathcal{A}}_-$ or $\hat{\tilde{\mathcal{A}}}_-$:

$$\hat{\mathcal{A}}_-\hat{\tilde{\mathcal{A}}}_-|0,\tilde{0}> = \left(\hat{\mathcal{A}}_-|0>\right)\otimes\left(\hat{\tilde{\mathcal{A}}}_-|\tilde{0}>\right) = 0 \quad (3.27)$$

By analogy with paper [8] where the canonical operators $\hat{a}^+$ and $\hat{a}$ intervene, for the TFD approach the following considerations can be formulated.

Nonlinear operators have the following structure

$$\hat{\mathcal{A}}_- = \hat{a}\, f(\hat{n}) \ , \quad \hat{\mathcal{A}}_+ = f(\hat{n})\, \hat{a}^+ \tag{3.28}$$

We will choose the nonlinearity function in the following manner

$$f(\hat{n}) = \frac{\prod_{j=1}^{q}(b_j - 1 + \hat{n})}{\prod_{i=1}^{p}(a_i - 1 + \hat{n})} \ , \quad f(\hat{n})|n> = \frac{\prod_{j=1}^{q}(b_j - 1 + n)}{\prod_{i=1}^{p}(a_i - 1 + n)}|n> \tag{3.29}$$

Also, we will perform the calculations using the DOOT technique (diagonal operator ordering technique), in which the main feature is that the operators $\hat{\mathcal{A}}_+$ and $\hat{\mathcal{A}}_-$ (and, consequently, $\hat{\tilde{\mathcal{A}}}_+$ and $\hat{\tilde{\mathcal{A}}}_-$) commute with each other inside the symbol #...#, in order to obtain directly ordered operator products. In addition, under the integrations and derivations and other algebraic operations the operators are treated as simple *c*-numbers [27]. Consequently, we have

$$\left[\hat{\mathcal{A}}_+, \hat{\mathcal{A}}_-\right] = 0 \ , \quad \left[\hat{\tilde{\mathcal{A}}}_+, \hat{\tilde{\mathcal{A}}}_-\right] = 0 \ , \quad \left[\hat{\mathcal{A}}_\pm, \hat{\tilde{\mathcal{A}}}_\pm\right] = 0 \ , \quad \left[\hat{\mathcal{A}}_\pm, \hat{\tilde{\mathcal{A}}}_\mp\right] = 0 \quad \text{(commut)}$$

In the next we use the abbreviated notation for the coefficients: $\boldsymbol{a} \equiv \{a_1, a_2, ..., a_p\} \equiv \{a_i\}_1^p$ and so on, and $_pF_q(\boldsymbol{a}; \boldsymbol{b}; x)$ is the generalized hypergeometric function.

The same rules will be applied also to operators acting in dual Hilbert tilde space $\tilde{\mathcal{H}}$.

To simplify writing formulas, everywhere possible, we will write the hypergeometric functions without mentioning the set of numbers $\boldsymbol{a}$ and $\boldsymbol{b}$ and indicating only their argument, e.g. $_pF_q(\boldsymbol{a}; \boldsymbol{b}; x) \equiv\, _pF_q(x)$. With this notation, the generalized hypergeometric function will be written as

$$_pF_q(\boldsymbol{a}; \boldsymbol{b}; x) \equiv\, _pF_q(x) = \sum_{n=0}^{\infty} \frac{\prod_{i=1}^{p}(a_i)_n}{\prod_{j=1}^{q}(b_j)_n} \frac{x^n}{n!} = \sum_{n=0}^{\infty} \frac{x^n}{\rho(n)} \tag{3.30}$$

One of the important advantages of using the DOOT technique (as well as the IWOP technique) is that it allows us to express various entities in terms of creation / annihilation operators $\hat{\mathcal{A}}_+$ and $\hat{\mathcal{A}}_-$. Apart from the thermal vacuum projector $|0(\beta)><0(\beta)|$ which is expressed as a function that depends on the normal ordered product of creation and annihilation operators $\#\hat{\mathcal{A}}_+\hat{\mathcal{A}}_-\#$ we will also use functions that have operators as arguments and, in the integral calculations, these operators are viewed as simple *c*-numbers, if they are framed by the symbols #...#.

So, essentially, expressions that obey the DOOT rules are enclosed within #...# symbols. Inside them, all operators (both those acting in the physical space and those corresponding to the tilde space) must be placed in a normal ordering: creation operators on the left, and annihilation operators on the right.

For zero temperature quantum field theory, the corresponding CSs, defined in the Barut-Girardello manner [17], have the following properties (where we used also the hermitian conjugates of these relations)

$$\hat{\mathcal{A}}_-|z>=z|z>, \qquad <z|\hat{\mathcal{A}}_-=0$$
$$<z|\hat{\mathcal{A}}_+=z^*<z|, \qquad \hat{\mathcal{A}}_+|z>=0 \qquad (3.31)$$
$$<z|\hat{\mathcal{A}}_+\hat{\mathcal{A}}_-|z>=|z|^2, \qquad <z|\hat{\mathcal{A}}_-\hat{\mathcal{A}}_+|z>=0$$

The results in the right column are due to the fact that $|z>$ is the eigenfunction of the operator $\hat{\mathcal{A}}_-$, and at the same time $<z|$ is the eigenfunction of the operator $\hat{\mathcal{A}}_+$.

Another useful relation in which the ordered product of operators intervenes is

$$<z|\mathcal{F}(\#\hat{\mathcal{A}}_+\hat{\mathcal{A}}_-\#)|z>=\mathcal{F}(|z|^2) \qquad (3.32)$$

which means that, if we have to calculate the expectation value, in the CSs representation, of a function that depends on the ordered product of $\#\hat{\mathcal{A}}_+\hat{\mathcal{A}}_-\#$ operators, the result is obtained simply by making the substitution $\#\hat{\mathcal{A}}_+\hat{\mathcal{A}}_-\# \to |z|^2$.

The generalized thermal raising and lowering operators (i.e. operators which depend on the temperature $T$ or $\beta$ are defined using the Bogoliubov transformation. Then, for a system with boson particles, the temperature dependent ladder operators are

$$\hat{\mathcal{A}}_-(\beta)=\cosh\theta(\beta)\hat{\mathcal{A}}_- - \sinh\theta(\beta)\hat{\tilde{\mathcal{A}}}_+$$
$$\hat{\mathcal{A}}_+(\beta)=\cosh\theta(\beta)\hat{\mathcal{A}}_+ - \sinh\theta(\beta)\hat{\tilde{\mathcal{A}}}_- \qquad (3.33)$$

The set of operators $\hat{\mathcal{A}}_+(\beta)$ and $\hat{\mathcal{A}}_-(\beta)$ which acts in the Hilbert physical and tilde spaces satisfies the same type of relations as $\hat{\mathcal{A}}_+$ and $\hat{\mathcal{A}}_-$. Their action on the whole vector basis are

$$\hat{\mathcal{A}}_-(\beta)|n,\tilde{n}>=\cosh\theta(\beta)\sqrt{e(n)}|n-1,\tilde{n}> - \sinh\theta(\beta)\sqrt{e(n+1)}|n+1,\tilde{n}>$$
$$\hat{\mathcal{A}}_+(\beta)|n,\tilde{n}>=\cosh\theta(\beta)\sqrt{e(n+1)}|n+1,\tilde{n}> - \sinh\theta(\beta)\sqrt{e(n)}|n-1,\tilde{n}> \qquad (3.34)$$

and similarly for the tilde operators.

In the frame of the DOOT technique, the creation and the annihilation operators are commutable, i.e we have

$$\#[\hat{\mathcal{A}}_-,\hat{\mathcal{A}}_+]\#=\#[\hat{\tilde{\mathcal{A}}}_-,\hat{\tilde{\mathcal{A}}}_+]\#=0, \quad \#[\hat{\mathcal{A}}_-,\hat{\tilde{\mathcal{A}}}_+]\#=\#[\hat{\tilde{\mathcal{A}}}_-,\hat{\mathcal{A}}_+]\#=0 \qquad (3.35)$$

This property is also transmitted to temperature dependent operators

$$\#[\hat{\mathcal{A}}_-(\beta),\hat{\mathcal{A}}_+(\beta)]\#=\#[\hat{\tilde{\mathcal{A}}}_-(\beta),\hat{\tilde{\mathcal{A}}}_+(\beta)]\#=0$$
$$\#[\hat{\mathcal{A}}_-(\beta),\hat{\tilde{\mathcal{A}}}_+(\beta)]\#=\#[\hat{\tilde{\mathcal{A}}}_-(\beta),\hat{\mathcal{A}}_+(\beta)]\#=0 \qquad (3.36)$$

As a consequence of the above properties, the thermal vacuum state $|0(\beta)>$ is annihilated by either $\hat{\mathcal{A}}_-(\beta)$ or $\hat{\tilde{\mathcal{A}}}_-(\beta)$, i.e.

$$\hat{\mathcal{A}}_-(\beta)|0(\beta)>=\hat{\tilde{\mathcal{A}}}_-(\beta)|0(\beta)>=|0> \qquad (3.37)$$

This can be showed as follows, e.g.:

$$\hat{\mathcal{A}}_-(\beta)|0(\beta)> = \left[\cosh\theta(\beta)\hat{\mathcal{A}}_- - \sinh\theta(\beta)\hat{\mathcal{A}}_+\right]|0> \otimes |\tilde{0}> =$$
$$= \cosh\theta(\beta)\left[\hat{\mathcal{A}}_-|0>\right] \otimes |\tilde{0}> - \sinh\theta(\beta)\left[\hat{\mathcal{A}}_+|0>\right] \otimes |\tilde{0}> = |0,\tilde{0}> \quad (3.38)$$

The second term in the second row cancels out because the operator $\hat{\mathcal{A}}_+$ has the bra vectors $<...|$ as eigenvectors and not the ket vectors $|...>$.

At the end of this section let us we calculate the expected value of the ordered normal product of thermal operators $\hat{\mathcal{A}}_+(\beta)\hat{\mathcal{A}}_-(\beta)$ in the thermal vacuum state $|0(\beta)>$. We have, successively

$$<0(\beta)|\hat{\mathcal{A}}_+(\beta)\hat{\mathcal{A}}_-(\beta)|0(\beta)> = \sum_n \sqrt{\frac{e^{-\beta E_n}}{Z(\beta)}} <n,\tilde{n}|\hat{\mathcal{A}}_+(\beta)\hat{\mathcal{A}}_-(\beta)\sum_m \sqrt{\frac{e^{-\beta E_m}}{Z(\beta)}}|m,\tilde{m}> =$$
$$= \sum_n \sqrt{\frac{e^{-\beta E_n}}{Z(\beta)}} \sum_m \sqrt{\frac{e^{-\beta E_m}}{Z(\beta)}} <n,\tilde{n}|\hat{\mathcal{A}}_+(\beta)\hat{\mathcal{A}}_-(\beta)|m,\tilde{m}> \quad (3.39)$$

$$<n,\tilde{n}|\hat{\mathcal{A}}_+(\beta)\hat{\mathcal{A}}_-(\beta)|m,\tilde{m}> =$$
$$= \left(<n|\left[\cosh\theta(\beta)\hat{\mathcal{A}}_+ - \sinh\theta(\beta)\hat{\mathcal{A}}_-\right]\left[\cosh\theta(\beta)\hat{\mathcal{A}}_- - \sinh\theta(\beta)\hat{\mathcal{A}}_+\right]|m>\right) \otimes <\tilde{n}|\tilde{m}> = \quad (3.40)$$
$$= \cosh^2\theta(\beta)e(n) + \sinh^2\theta(\beta)e(n+1)$$

$$<0(\beta)|\hat{\mathcal{A}}_+(\beta)\hat{\mathcal{A}}_-(\beta)|0(\beta)> = \frac{1}{Z(\beta)}\sum_n e^{-\beta E_n}\left[\cosh^2\theta(\beta)e(n) + \sinh^2\theta(\beta)e(n+1)\right] =$$
$$= \cosh^2\theta(\beta)\frac{1}{Z(\beta)}\sum_n e^{-\beta\hbar\omega e(n)}e(n) + \sinh^2\theta(\beta)\frac{1}{Z(\beta)}\sum_n e^{-\beta\beta\hbar\omega e(n)}e(n+1) \quad (3.41)$$

The final result is then, see Eq. (3.13)

$$<0(\beta)|\hat{\mathcal{A}}_+(\beta)\hat{\mathcal{A}}_-(\beta)|0(\beta)> = -\cosh^2\theta(\beta)\frac{\partial}{\partial(\beta\hbar\omega)}\ln Z(\beta) + \sinh^2\theta(\beta)\frac{1}{Z(\beta)}\sum_n e^{-\beta\hbar\omega e(n)}e(n+1) \quad (3.42)$$

The second term can only be evaluated if the relationship between $e(n+1)$ and $e(n)$ is known, which is not the case for the generalized energy spectrum, Eq. (3.4).

But, for example, for systems with a linear energy spectrum, whose distance between energy levels is constant, of the form

$$E_n = \hbar\omega e(n) = \hbar\omega\left(n + \frac{E_0}{\hbar\omega}\right), \quad e(n+1) = \left(n+1+\frac{E_0}{\hbar\omega}\right) = e(n) + 1 \quad (3.43)$$

and the calculation can continue.

$$\frac{1}{Z(\beta)}\sum_n e^{-\beta\hbar\omega e(n)}e(n+1) = \frac{1}{Z(\beta)}\sum_n e^{-\beta\hbar\omega e(n)}e(n) + 1 \quad (3.44)$$

Finally, we will obtain

$$<0(\beta)|\hat{\mathcal{A}}_+(\beta)\hat{\mathcal{A}}_-(\beta)|0(\beta)> = \sinh^2\theta(\beta) - \left[\cosh^2\theta(\beta) + \sinh^2\theta(\beta)\right]\frac{\partial}{\partial(\beta\hbar\omega)}\ln Z(\beta) =$$
$$= \sinh^2\theta(\beta) + \cosh 2\theta(\beta)\frac{1}{\hbar\omega}U(\beta) \quad (3.45)$$

Using the limits $\lim_{\theta \to 0} \hat{\mathcal{A}}_+(\beta) = \hat{\mathcal{A}}_+$ and $\lim_{\theta \to 0} \hat{\mathcal{A}}_-(\beta) = \hat{\mathcal{A}}_-$ we will obtain the previous result, Eq. (3.13).

At the end of this section, let's see how (or if) the generalized temperature dependent vacuum state $|0(\beta)>_{gen}$ can be defined by means of the thermal dependent operators $\hat{\mathcal{A}}_+(\beta)$ and $\hat{\mathcal{A}}_-(\beta)$. We will denote this new temperature dependent vacuum state with $|0(\beta;g(\theta))>_{gen} \equiv |0(\beta;g)>_{gen}$. Here $g(\theta)$ is some generalized deformation factor whose expression we have to determine. For now, let's write it in the form $[g(\theta)]^{\hat{n}}$.

We assume that the generalized vacuum temperature dependent state is defined as follows:

$$|0(\beta;g)>_{gen} \equiv \sqrt{\frac{e^{-\beta \hat{\mathcal{H}}}}{Z(\beta)}} [g(\theta)]^{\hat{n}} \sum_n |n,\tilde{n}> = \sum_n \sqrt{\frac{e^{-\beta E_n}}{Z(\beta)}} [g(\theta)]^n |n,\tilde{n}> \qquad (3.46)$$

Let's express the vector $|n,\tilde{n}>$ as a result of the action of temperature dependent operators on the double vacuum state

$$\left[\hat{\mathcal{A}}_+(\beta)\hat{\tilde{\mathcal{A}}}_+(\beta)\right]^n |0,\tilde{0}> = (\cosh^2\theta)^n {}_p\rho_q(n)|n,\tilde{n}> \qquad (3.47)$$

Then we obtain

$$|0(\beta;g)>_{gen} = \sum_n \sqrt{\frac{e^{-\beta E_n}}{Z(\beta)}} \frac{1}{{}_p\rho_q(n)} \left[\frac{g(\theta)}{\cosh^2\theta} \hat{\mathcal{A}}_+(\beta)\hat{\tilde{\mathcal{A}}}_+(\beta)\right]^n |0,\tilde{0}> =$$

$$= \sum_n \sqrt{\frac{e^{-\beta E_n}}{Z(\beta)}} \frac{\left(\hat{\mathcal{A}}_+\hat{\tilde{\mathcal{A}}}_+\right)^n}{{}_p\rho_q(n)} |0,\tilde{0}> \qquad (3.48)$$

From here it follows that $g(\theta)=1$, meaning $|0(\beta;g)>_{gen}$ does not depend on the generalized deformation factor, i.e. the generalized $|0(\beta;g)>_{gen} \equiv |0(\beta)>_{gen}$:

$$|0(\beta)>_{gen} = \sum_n \sqrt{\frac{e^{-\beta E_n}}{Z(\beta)}} \frac{1}{{}_p\rho_q(n)} \left[\frac{\hat{\mathcal{A}}_+(\beta)\hat{\tilde{\mathcal{A}}}_+(\beta)}{\cosh^2\theta}\right]^n |0,\tilde{0}> = \sum_n \sqrt{\frac{e^{-\beta E_n}}{Z(\beta)}} \frac{\left(\hat{\mathcal{A}}_+\hat{\tilde{\mathcal{A}}}_+\right)^n}{{}_p\rho_q(n)} |0,\tilde{0}> \quad (3.49)$$

In the limit for ${}_pf_q(n)=1$, the generalized vacuum temperature dependent state becomes equal to the usual vacuum temperature dependent state, that for a system of free bosons:

$$\lim_{{}_pf_q(n)=1} |0(\beta)>_{gen} \equiv \lim_{\substack{p=q \\ \{a_i\}=\{b_j\}}} |0(\beta)>_{gen} = |0(\beta)> = \frac{1}{\cosh\theta(\beta)} e^{\tanh\theta(\beta)\hat{a}^+\hat{\tilde{a}}^+} |0,\tilde{0}> \qquad (3.50)$$

## 4. Generalized coherent states in the frame of TFD

### 4.1 Pure states

Because the thermal coherent states (TCSs) can be defined as a natural extension of the corresponding usual CSs, these can be defined in two ways: a) in the sense of Barut and Girardello, i.e. as eigenvalues of the annihilation operator; b) in the sense of Klauder and Perelomov, i.e. as the result of the action of the displacement operator on the vacuum state vector. Generally, a coherent state (CS) is labeled by a the complex variable, e.g. $z = |z|\exp(i\varphi)$, with $|z| \leq \infty$ and $\varphi \in (0, 2\pi)$.

### a) Definition according to the Barut and Girardello procedure

Usually, the CSs in the Barut-Girradello manner (BG-CSs) are defined as the eigenvector of the annihilation operator, the eigenequation being [17].

$$\hat{\mathcal{A}}_- | z >_{BG} = z | z >_{BG} \tag{4.1}$$

Simultaneously, CSs $|z>_{BG}$ are not eigenvectors of the creation operator $\hat{\mathcal{A}}_+$. Therefore, $\hat{\mathcal{A}}_+ | z >_{BG} = 0 \cdot | z >_{BG}$. The same is also satisfied by complex conjugate relations:

$$_{BG}<z|\hat{\mathcal{A}}_+ = z^* {}_{BG}<z| \quad , \quad _{BG}<z|\hat{\mathcal{A}}_- = 0 \cdot {}_{BG}<z| \tag{4.2}$$

Indeed, the operator $\hat{\mathcal{A}}_+$ has the bra vector $_{BG}<z|$ (with argument $z^*$) as their eigenvector and $z^*$ as their eigenvalue.

Now, if we apply the temperature dependent operator on the usual CSs, we will be able to write, successively

$$\hat{\mathcal{A}}_-(\beta) | z; \beta >_{BG} = \#\left[\cosh\theta(\beta)\hat{\mathcal{A}}_- - \sinh\theta(\beta)\hat{\mathcal{A}}_+\right]\# | z; \beta >_{BG} =$$
$$= \left[-\sinh\theta(\beta)\hat{\mathcal{A}}_+\right] | z; \beta >_{BG} + \left[\cosh\theta(\beta)\hat{\mathcal{A}}_-\right] | z; \beta >_{BG} = \tag{4.3}$$
$$= \left[\cosh\theta(\beta)\hat{\mathcal{A}}_-\right] | z; \beta >_{BG} = z\cosh\theta(\beta) | z; \beta >_{BG}$$

So, finally, we got

$$\hat{\mathcal{A}}_-(\beta) | z; \beta >_{BG} = z\cosh\theta(\beta) | z; \beta >_{BG} \tag{4.4}$$

Their complex conjugate partner is then

$$_{BG}<z; \beta |\hat{\mathcal{A}}_+(\beta) = z^* \cosh\theta(\beta) {}_{BG}<z; \beta | \tag{4.5}$$

and their normal ordered product leads to

$$_{BG}<z; \beta |\#\hat{\mathcal{A}}_+(\beta)\hat{\mathcal{A}}_-(\beta)\#| z; \beta >_{BG} = |z|^2 \cosh^2\theta(\beta) \tag{4.6}$$

as well as

$$_{BG}<z; \beta | \mathcal{F}\left(\#\hat{\mathcal{A}}_+(\beta)\hat{\mathcal{A}}_-(\beta)\#\right)| z; \beta >_{BG} = \mathcal{F}\left(|z|^2 \cosh^2\theta(\beta)\right) \tag{4.7}$$

Absolutely similar relations can be written for the CSs of tilde Hilbert space.

The whole coherent states (for the total space, the tensor product $\mathsf{H} \otimes \tilde{\mathsf{H}}$) can be defined as a tensor product of the coherent states for the individual spaces (original $\otimes$ tilde):

$$| z, \tilde{\sigma}; \beta >_{BG} \equiv | z; \beta >_{BG} \otimes | \tilde{\sigma}; \beta >_{BG} \tag{4.8}$$

where, for the tilde space we used the complex variable $\tilde{\sigma}$, to avoid confusion.

In fact, the whole CSs can be considered as two-mode CSs.

The whole CSs are the eigenvectors of the individual temperature dependent operators $\hat{\mathcal{A}}_-(\beta)$ and $\tilde{\hat{\mathcal{A}}}_-(\beta)$:

$$\hat{\mathcal{A}}_-(\beta) | z, \tilde{\sigma}; \beta >_{BG} = z\cosh\theta(\beta) | z, \tilde{\sigma}; \beta >_{BG} \tag{4.9}$$

$$_{BG}<z,\tilde{\sigma};\beta|\hat{\mathcal{A}}_+(\beta)=z^*\cosh\theta(\beta)\,_{BG}<z,\tilde{\sigma};\beta| \tag{4.10}$$

$$_{BG}<z,\tilde{\sigma};\beta|\hat{\mathcal{A}}_+(\beta)\hat{\mathcal{A}}_-(\beta)|z,\tilde{\sigma};\beta>_{BG}=|z|^2\cosh^2\theta(\beta) \tag{4.11}$$

$$\hat{\tilde{\mathcal{A}}}_-(\beta)|z,\tilde{\sigma};\beta>_{BG}=\tilde{\sigma}\cosh\theta(\beta)|z,\tilde{\sigma};\beta>_{BG} \tag{4.12}$$

$$_{BG}<z,\tilde{\sigma};\beta|\hat{\tilde{\mathcal{A}}}_+(\beta)=\tilde{\sigma}^*\cosh\theta(\beta)\,_{BG}<z,\tilde{\sigma};\beta| \tag{4.13}$$

$$\hat{\mathcal{A}}_-(\beta)\hat{\tilde{\mathcal{A}}}_-(\beta)|z,\tilde{\sigma};\beta>_{BG}=z\tilde{\sigma}\cosh^2\theta(\beta)|z,\tilde{\sigma};\beta>_{BG} \tag{4.14}$$

$$_{BG}<z,\tilde{\sigma};\beta|\hat{\mathcal{A}}_+(\beta)\hat{\tilde{\mathcal{A}}}_+(\beta)=z^*\tilde{\sigma}^*\cosh^2\theta(\beta)\,_{BG}<z,\tilde{\sigma};\beta| \tag{4.15}$$

Now, for the construction of TFD CSs we apply the procedure of Man and Revzen [3], which consists in the natural extension of the definition to thermal coherent states (TCSs). They defined coherent states in the sense of Perelomov, that is, as the result of the action of the displacement operator on the vacuum state vector. Since in TFD, the "normal" and "tilde" operators commute each with other, besides the usual Hilbert space of absolute zero temperature $\mathcal{H}$, the so-called tilde space $\tilde{\mathcal{H}}$ also intervenes, in the form of a direct product, i. e. $\mathcal{H}\otimes\tilde{\mathcal{H}}$, the displacement operator will be written in the form, with respect to the DOOT rules

$$\hat{\mathcal{D}}(z,\tilde{\sigma};\beta)\equiv\hat{\mathcal{D}}(z;\beta)\otimes\hat{\tilde{\mathcal{D}}}(\tilde{\sigma};\beta) \tag{4.16}$$

We have previously seen that the generalized CSs can also be obtained by applying the generalized hypergeometric function $_pF_q(\boldsymbol{a};\boldsymbol{b};z\hat{\mathcal{A}}_+)$ on the vacuum ket vector $|0>$, see, Eq. (3.14). This idea appeared for the first time, for a particular case of pseudoharmonic oscillator, in Mojaveri and Dehghani's paper [30].

In order not to complicate the writing of formulas, where no confusion can occur, we will henceforth omit writing $\boldsymbol{a};\boldsymbol{b}$ in the expressions of generalized hypergeometric functions $_pF_q(...)$, writing only the argument, but will not omit that these depend on the sets of numbers $\boldsymbol{a}$ and $\boldsymbol{b}$.

If to this idea is also added the fact that, since $\hat{\mathcal{A}}_-|0>=|0>$ any function with the annihilation operator as argument, therefore also $_pF_q(;z^*\hat{\mathcal{A}}_-)|0>=|0>$ leaves unchanged.

Then it is obvious that a generalized displacement operator can be constructed, similar to the "usual" one $D(z)=e^{-\frac{1}{2}|z|^2}e^{za^+-z^*a}$.

In this sense the normalized generalized hypergeometric displacement operator is

$$\hat{\mathcal{D}}(z)=\frac{1}{\sqrt{_pF_q(;|z|^2)}}\,_pF_q(;z\hat{\mathcal{A}}_+) \tag{4.17}$$

In zero temperature quantum field theory the generalized displacement operator is defined similarly

$$\hat{\mathcal{D}}(z;\beta)=\frac{1}{\sqrt{_pN_q(z;\beta)}}\,_pF_q(;z\hat{\mathcal{A}}_+(\beta)) \tag{4.18}$$

As in the zero temperature case, coherent states are the result of the action of the displacement operator on the whole vacuum state.

$$|z, \tilde{\sigma}; \beta>_{BG} = \frac{1}{\sqrt{{}_pN_q(z, \tilde{\sigma}; \beta)}} \hat{\mathcal{D}}(z, \tilde{\sigma}; \beta)|0, \tilde{0}>=$$

$$= \frac{1}{\sqrt{{}_pN_q(z, \tilde{\sigma}; \beta)}} \left[ \frac{1}{\sqrt{{}_pN_q(z; \beta)}} \hat{\mathcal{D}}(z; \beta)|0> \right] \otimes \left[ \frac{1}{\sqrt{{}_pN_q(\tilde{\sigma}; \beta)}} \hat{\tilde{\mathcal{D}}}(\tilde{\sigma}; \beta)|\tilde{0}> \right] \equiv \quad (4.19)$$

$$\equiv \frac{1}{\sqrt{{}_pN_q(z, \tilde{\sigma}; \beta)}} |z; \beta>_{BG} \otimes |\tilde{\sigma}; \beta>_{BG}$$

Let's perform the calculations for the BG-CSs of original Hilbert space.

$$|z; \beta>_{BG} = \frac{1}{\sqrt{{}_pN_q(z; \beta)}} \hat{\mathcal{D}}(z; \beta)|0> = \frac{1}{\sqrt{{}_pN_q(z; \beta)}} {}_pF_q\left(; z\hat{\mathcal{A}}_+(\beta)\right)|0>=$$

$$= \frac{1}{\sqrt{{}_pN_q(z; \beta)}} \sum_{n=0}^{\infty} \frac{z^n}{{}_p\rho_q^{BG}(n)} \left[\hat{\mathcal{A}}_+(\beta)\right]^n |0> \quad (4.20)$$

$$\left[\hat{\mathcal{A}}_+(\beta)\right]^n |0> = \left[z\cosh\theta(\beta)\hat{\mathcal{A}}_+ - z\sinh\theta(\beta)\hat{\mathcal{A}}_-\right]^n |0> =$$

$$= \sum_{k=0}^{n} \binom{n}{k} \left[z\cosh\theta(\beta)\hat{\mathcal{A}}_+\right]^{n-k} \left[-z\sinh\theta(\beta)\hat{\mathcal{A}}_-\right]^k |0> \quad (4.21)$$

Since $\hat{\mathcal{A}}_-|0>=|0>$, the equality remains valid only for $k=0$, and we obtain

$$\left[\hat{\mathcal{A}}_+(\beta)\right]^n |0> = \left[z\cosh\theta(\beta)\hat{\mathcal{A}}_+\right]^n |0> \quad (4.22)$$

Finally, the BG-CSs for physical system becomes

$$|z; \beta>_{BG} = \frac{1}{\sqrt{{}_pN_q(z; \beta)}} \sum_{n=0}^{\infty} \frac{\prod_{i=1}^{p}(a_i)_n}{\prod_{j=1}^{q}(b_j)_n} \frac{[z\cosh\theta(\beta)]^n}{n!} \left(\hat{\mathcal{A}}_+\right)^n |0>=$$

$$= \frac{1}{\sqrt{{}_pN_q(z; \beta)}} {}_pF_q\left(\boldsymbol{a}; \boldsymbol{b}; z\cosh\theta(\beta)\hat{\mathcal{A}}_+\right)|0> \quad (4.23)$$

or, equivalently

$$|z; \beta>_{BG} = \frac{1}{\sqrt{{}_pN_q(z; \beta)}} \sum_{n=0}^{\infty} \frac{[z\cosh\theta(\beta)]^n}{\sqrt{{}_p\rho_q^{BG}(n)}} |n> \quad (4.24)$$

The normalization function is

$$_pN_q(z; \beta) = \sum_{n=0}^{\infty} \frac{\left[|z|^2 \cosh^2\theta(\beta)\right]^n}{{}_p\rho_q^{BG}(n)} = {}_pF_q\left(; |z|^2 \cosh^2\theta(\beta)\right) \quad (4.25)$$

so that the final expression of these BG-CSs is

$$|z;\beta>_{BG} = \frac{1}{\sqrt{{}_pF_q(;|z|^2\cosh^2\theta(\beta))}} \sum_{n=0}^{\infty} \frac{[z\cosh\theta(\beta)]^n}{\sqrt{{}_p\rho_q^{BG}(n)}} |n> \qquad (4.26)$$

Since the variables $z$ and $\tilde{\sigma}$ are symmetric (i.e. "have the same rights"), each set of coherent states must satisfy the completeness relation (i.e., the decomposition of the unit operator). For example, for the variable $z$ this relation is

$$\int d\mu_{BG}(z;\beta)|z;\beta>_{BG}{}_{BG}<z;\beta| = I \qquad (4.27)$$

We suppose that the integration measure has the following structure

$$d\mu_{BG}(z;\beta) = \frac{d\varphi}{2\pi} d(|z|^2) h(|z|^2) \qquad (4.28)$$

where the weight function $h(|z|^2)$ must be determined. So, we obtain

$$\sum_{n=0}^{\infty} \frac{[\cosh\theta(\beta)]^n}{\sqrt{{}_p\rho_q^{BG}(n)}} |n> \sum_{m=0}^{\infty} \frac{[\cosh\theta(\beta)]^m}{\sqrt{{}_p\rho_q^{BG}(m)}} <m| \times$$

$$\times \int_0^{\infty} d(|z|^2 \cosh^2\theta(\beta)) \frac{h(|z|^2)}{{}_pF_q(;|z|^2\cosh^2\theta(\beta))} \int_0^{2\pi} \frac{d\varphi}{2\pi} z^n(z^*)^m = 1 \qquad (4.29)$$

The angular integral is equal to $(|z|^2)^n \delta_{nm}$ so that we have

$$\sum_{n=0}^{\infty} \frac{|n><n|}{{}_p\rho_q^{BG}(n)} \int_0^{\infty} d(|z|^2\cosh^2\theta(\beta)) \frac{h(|z|^2)}{{}_pF_q(;|z|^2\cosh^2\theta(\beta))} (|z|^2\cosh^2\theta(\beta))^n = 1 \qquad (4.30)$$

In order to satisfy the closure relation of the Fock vectors $\sum_{n=0}^{\infty} |n><n| = 1$ we must have the following equality satisfied

$$\int_0^{\infty} d(|z|^2\cosh^2\theta(\beta)) \frac{h(|z|^2)}{{}_pF_q(;|z|^2\cosh^2\theta(\beta))} [|z|^2\cosh^2\theta(\beta)]^n = {}_p\rho_q^{BG}(n) \qquad (4.31)$$

After changing the index, $n = s-1$ and writing explicitly the structure constant, we will arrive at a Stieltjes-type moment problem:

$$\int_0^{\infty} d(|z|^2\cosh^2\theta(\beta)) \frac{h(|z|^2)}{{}_pF_q(;|z|^2\cosh^2\theta(\beta))} [|z|^2\cosh^2\theta(\beta)]^{s-1} = \Gamma(\boldsymbol{a}/\boldsymbol{b})\Gamma(s) \frac{\prod_{j=1}^{q}\Gamma(b_j-1+s)}{\prod_{i=1}^{p}\Gamma(a_i-1+s)} \qquad (4.32)$$

The solution of this problem is expressed through Meijer $G$-function [31]

$$\frac{h(|z|^2)}{{}_pF_q(;|z|^2\cosh^2\theta(\beta))} = \Gamma(\boldsymbol{a}/\boldsymbol{b}) G_{p,q+1}^{q+1,0}\left(|z|^2\cosh^2\theta(\beta) \middle| \begin{array}{cc} /\,; & \boldsymbol{a-1} \\ 0,\,\boldsymbol{b-1}\,; & / \end{array}\right) \qquad (4.33)$$

Then, the final expression of the integration measure is

$$d\mu_{BG}(z;\beta) = \Gamma(\boldsymbol{a}/\boldsymbol{b}) \frac{d\varphi}{2\pi} d(|z|^2\cosh^2\theta(\beta)) \times$$

$$\times {}_pF_q(;|z|^2\cosh^2\theta(\beta)) G_{p,q+1}^{q+1,0}\left(|z|^2\cosh^2\theta(\beta) \middle| \begin{array}{cc} /\,; & \boldsymbol{a-1} \\ 0,\,\boldsymbol{b-1}\,; & / \end{array}\right) \qquad (4.34)$$

Using the classical and generalized Meijer's definite integral from one $G$ function [31]

$$\int_0^\infty dt\, t^{\alpha-1} G_{p,q}^{m,n}\left(tz \left|\begin{array}{c} a_1,a_2,...,a_n \; ; \; a_{n+1},a_{n+2},...,a_p \\ b_1,b_2,...,b_m \; ; \; b_{m+1},b_{m+2},...,b_q \end{array}\right.\right) = \frac{\prod_{j=1}^{m}\Gamma(\alpha+b_j)\prod_{i=1}^{n}\Gamma(1-\alpha-a_i)}{\prod_{i=n+1}^{p}\Gamma(\alpha+a_i)\prod_{j=m+1}^{q}\Gamma(1-\alpha-b_j)} z^{-\alpha} \quad (4.35)$$

it is not difficult to show, as being expected, that the integral od integration measure is equal to unity: $\int d\mu_{BG}(z;\beta) = 1$.

For the variable $\tilde{\sigma}$ the expression is similar.

The whole CSs corresponding to the whole Hilbert space $\mathsf{H}_{tot} = \mathsf{H} \otimes \tilde{\mathsf{H}}$ is

$$|z, \tilde{\sigma}; \beta>_{BG} = \frac{1}{\sqrt{{}_pN_q(z, \tilde{\sigma}; \beta)}} |z; \beta>_{BG} \otimes |\tilde{\sigma}; \beta>_{BG} =$$

$$= \left[\frac{{}_pF_q\left(; z\cosh\theta(\beta)\hat{\mathcal{A}}_+\right)}{\sqrt{{}_pF_q\left(; |z|^2 \cosh^2\theta(\beta)\right)}}\right]\left[\frac{{}_pF_q\left(; \tilde{\sigma}\cosh\theta(\beta)\hat{\tilde{\mathcal{A}}}_+\right)}{\sqrt{{}_pF_q\left(; |\tilde{\sigma}|^2 \cosh^2\theta(\beta)\right)}}\right]|0,\tilde{0}> \quad (4.36)$$

At the same time, their complex conjugate counterpart is

$${}_{BG}<z, \tilde{\sigma}; \beta| = <0,\tilde{0}|\left[\frac{{}_pF_q\left(; z\cosh\theta(\beta)\hat{\mathcal{A}}_-\right)}{\sqrt{{}_pF_q\left(; |z|^2 \cosh^2\theta(\beta)\right)}}\right]\left[\frac{{}_pF_q\left(; \tilde{\sigma}\cosh\theta(\beta)\hat{\tilde{\mathcal{A}}}_-\right)}{\sqrt{{}_pF_q\left(; |\tilde{\sigma}|^2 \cosh^2\theta(\beta)\right)}}\right] \quad (4.37)$$

Since the CSs depending on the variables $z$ and $\tilde{\sigma}$ have been previously normalized, the whole normalization factor is obtained from the condition: ${}_{BG}<z, \tilde{\sigma}; \beta | z, \tilde{\sigma}; \beta>_{BG} = 1$

$${}_{BG}<z, \tilde{\sigma}; \beta | z, \tilde{\sigma}; \beta> = \frac{1}{{}_pN_q(z, \tilde{\sigma}; \beta)} \times$$

$$\times <0,\tilde{0}|\left[\frac{{}_pF_q\left(; z^*\cosh\theta(\beta)\hat{\mathcal{A}}_-\right)}{\sqrt{{}_pF_q\left(; |z|^2 \cosh^2\theta(\beta)\right)}}\right]\left[\frac{{}_pF_q\left(; \tilde{\sigma}^*\cosh\theta(\beta)\hat{\tilde{\mathcal{A}}}_-\right)}{\sqrt{{}_pF_q\left(; |\tilde{\sigma}|^2 \cosh^2\theta(\beta)\right)}}\right] \times \quad (4.38)$$

$$\times \left[\frac{{}_pF_q\left(; z^*\cosh\theta(\beta)\hat{\mathcal{A}}_+\right)}{\sqrt{{}_pF_q\left(; |z|^2 \cosh^2\theta(\beta)\right)}}\right]\left[\frac{{}_pF_q\left(; \tilde{\sigma}^*\cosh\theta(\beta)\hat{\tilde{\mathcal{A}}}_+\right)}{\sqrt{{}_pF_q\left(; |\tilde{\sigma}|^2 \cosh^2\theta(\beta)\right)}}\right]|0,\tilde{0}> = 1$$

After the appropriate substitutions, we obtain

$${}_{BG}<z, \tilde{\sigma}; \beta | z, \tilde{\sigma}; \beta>_{BG} = \frac{1}{{}_pN_q(z, \tilde{\sigma}; \beta)}(<z;\beta|\otimes<\tilde{\sigma};\beta|)(|z;\beta>\otimes|\tilde{\sigma};\beta>) =$$

$$= \frac{1}{{}_pN_q(z, \tilde{\sigma}; \beta)}(<z;\beta|z;\beta>)\otimes(<\tilde{\sigma};\beta|\tilde{\sigma};\beta>) = 1 \quad (4.39)$$

Consequently, $_pN_q(z, \tilde{\sigma}; \beta) = 1$, and the normalized whole or global CSs are

$$|z, \tilde{\sigma}; \beta>_{BG} = \left[\frac{_pF_q\left(; z\cosh\theta(\beta)\hat{\mathcal{A}}_+\right)}{\sqrt{_pF_q(; |z|^2 \cosh^2\theta(\beta))}}\right]\left[\frac{_pF_q\left(; \tilde{\sigma}\cosh\theta(\beta)\hat{\tilde{\mathcal{A}}}_+\right)}{\sqrt{_pF_q(; |\tilde{\sigma}|^2 \cosh^2\theta(\beta))}}\right]|0, \tilde{0}>=$$

$$= \left[\frac{_pF_q\left(; z\cosh\theta(\beta)\hat{\mathcal{A}}_+\right)}{\sqrt{_pF_q(; |z|^2 \cosh^2\theta(\beta))}}\right]|0> \otimes \left[\frac{_pF_q\left(; \tilde{\sigma}\cosh\theta(\beta)\hat{\tilde{\mathcal{A}}}_+\right)}{\sqrt{_pF_q(; |\tilde{\sigma}|^2 \cosh^2\theta(\beta))}}\right]|0> \quad (4.40)$$

or in an alternative manner:

$$|z, \tilde{\sigma}; \beta>_{BG} = \frac{1}{\sqrt{_pF_q(; |z|^2 \cosh^2\theta(\beta))}}\sum_{n=0}^{\infty}\frac{[z\cosh\theta(\beta)]^n}{\sqrt{_p\rho_q^{BG}(n)}}|n> \otimes$$

$$\otimes \frac{1}{\sqrt{_pF_q(; |\tilde{\sigma}|^2 \cosh^2\theta(\beta))}}\sum_{n=0}^{\infty}\frac{[\tilde{\sigma}\cosh\theta(\beta)]^n}{\sqrt{_p\rho_q^{BG}(n)}}|\tilde{n}> \quad (4.41)$$

Let us now verify the validity of this relationship, referring to the case of a system of harmonic oscillators, for which $p = q = 0$, when $\hat{\mathcal{A}}_- = \hat{a}$, $\hat{\mathcal{A}}_+ = \hat{a}^+$ and $_pF_q(/; /; x) = e^x$. The reason is that for the one-dimensional harmonic oscillator the Barut-Girardello coherent states have the same expression as the Klauder-Perelomov coherent states (KP-CSs). Therefore we will not write the indices *BG* and *KP*, respectively.

$$|z, \tilde{\sigma}; \beta>= \left[e^{-\frac{1}{2}|z|^2\cosh^2\theta(\beta)+z\cosh\theta(\beta)\hat{a}^+}\right]|0> \otimes \left[e^{-\frac{1}{2}|\tilde{\sigma}|^2\cosh^2\theta(\beta)+\tilde{\sigma}\cosh\theta(\beta)\hat{\tilde{a}}^+}\right]|0>$$

$$<z, \tilde{\sigma}; \beta|=<0|\left[e^{-\frac{1}{2}|z|^2\cosh^2\theta(\beta)+z\cosh\theta(\beta)\hat{a}}\right] \otimes <0|\left[e^{-\frac{1}{2}|\tilde{\sigma}|^2\cosh^2\theta(\beta)+\tilde{\sigma}\cosh\theta(\beta)\hat{\tilde{a}}}\right] \quad (4.42)$$

Then, the completeness relation of the CSs (the decomposition of the unity operator) is

$$\int d(z, \tilde{\sigma}; \beta)|z, \tilde{\sigma}; \beta><z, \tilde{\sigma}; \beta|= \int d(z; \beta)|z; \beta><z; \beta| \otimes \int d(\tilde{\sigma}; \beta)|\tilde{\sigma}; \beta><\tilde{\sigma}; \beta|=1 \quad (4.43)$$

Let's calculate the integral over the variable z.

$$\int d(z; \beta)|z; \beta><z; \beta|=$$

$$= \int_0^\infty \frac{d^2(|z|^2\cosh^2\theta(\beta))}{\pi}\left[e^{-\frac{1}{2}|z|^2\cosh^2\theta(\beta)+z\cosh\theta(\beta)\hat{a}^+}\right]|0><0|\left[e^{-\frac{1}{2}|z|^2\cosh^2\theta(\beta)+z\cosh\theta(\beta)\hat{a}}\right]= \quad (4.44)$$

$$= \int_0^\infty \frac{d^2(|z|^2\cosh^2\theta(\beta))}{\pi}:e^{-|z|^2\cosh^2\theta(\beta)+z\cosh\theta(\beta)\hat{a}^++z\cosh\theta(\beta)\hat{a}}:|0><0|=:e^{\hat{a}^+\hat{a}}\frac{1}{e^{\hat{a}^+\hat{a}}}:=1$$

In ordering the operators, we applied the rules of the IWOP technique, which uses the sign :...: instead of the #...# sign. We used the IWOP expression for the vacuum projector $|0><0|$, as well as the following integral in complex space [5]:

$$\int \frac{d^2z}{\pi}e^{A|z|^2+Bz+Cz^*} = -\frac{1}{A}e^{-\frac{BC}{A}} \quad (4.45)$$

This obtained expression is characteristic for a boson system associated with the one dimensional harmonic oscillator, because the quantum field for a harmonic oscillator is equivalent to a system of non-interacting bosons (e.g photons), where each oscillation mode correspond to a single boson. Since the energy of a one-dimensional harmonic oscillator is $E_n = \hbar\omega\left(n+\frac{1}{2}\right)$, partition function is $Z(\beta) = e^{-\beta\frac{\hbar\omega}{2}}\left(1-e^{-\beta\hbar\omega}\right)^{-1}$ and the expression of the thermal vacuum becomes

$$|0(\beta)\rangle = \sum_{n=0}^{\infty}\sqrt{\frac{e^{-\beta E_n}}{Z(\beta)}}|n,\tilde{n}\rangle = \sum_{n=0}^{\infty}\sqrt{\frac{e^{-\beta\hbar\omega\left(n+\frac{1}{2}\right)}}{Z(\beta)}}|n\rangle\otimes|\tilde{n}\rangle =$$

$$= \sqrt{1-e^{-\beta\hbar\omega}}\sum_{n=0}^{\infty}\left(e^{-\beta\frac{\hbar\omega}{2}}\right)^n\frac{(\hat{a}^+)^n}{\sqrt{n!}}|0\rangle\otimes\frac{(\tilde{\hat{a}}^+)^n}{\sqrt{n!}}|\tilde{0}\rangle = \sqrt{1-e^{-\beta\hbar\omega}}\sum_{n=0}^{\infty}\frac{1}{n!}\left(e^{-\beta\frac{\hbar\omega}{2}\hat{a}^+\tilde{\hat{a}}^+}\right)^n|0,\tilde{0}\rangle = \quad (4.46)$$

$$= \sqrt{1-e^{-\beta\hbar\omega}}\exp\left(e^{-\beta\frac{\hbar\omega}{2}\hat{a}^+\tilde{\hat{a}}^+}\right)|0,\tilde{0}\rangle = \frac{1}{\cosh\theta(\beta)}e^{\tanh\theta(\beta)\hat{a}^+\tilde{\hat{a}}^+}|0,\tilde{0}\rangle$$

So, finally we have another expression for the thermal vacuum of the system of free bosons, i.e. the formula for a single-mode thermal vacuum state of free bosons, in a compact, exponentially squeezed expression.

$$|0(\beta)\rangle = \frac{1}{\cosh\theta(\beta)}e^{\tanh\theta(\beta)\hat{a}^+\tilde{\hat{a}}^+}|0,\tilde{0}\rangle \quad (4.47)$$

**b) Definition according to the Klauder and Perelomov procedure**

We define the coherent states in the Klauder-Perelomov manner (KP-CSs) in TFD similar as there for zero temperature case, i.e. as the action of the exponential of creation operator on the vacuum state (the so called Klauder and Perelomov procedure, having in mind that $\hat{\mathcal{A}}_-|0\rangle = |0\rangle$). Recalling the usual definition

$$|z\rangle_{KP} = \frac{1}{\sqrt{N(|z|^2)}}\exp\left[z\hat{\mathcal{A}}_+\right]|0\rangle \quad (4.48)$$

in the TDF we can write

$$|z;\beta\rangle_{KP} = \frac{1}{\sqrt{N_{KP}(z;\beta)}}\exp\left[z\hat{\mathcal{A}}_+(\beta)\right]|0\rangle \quad (4.49)$$

Expanding the exponential in the power series and using the Newton binomial formula, we have

$$|z;\beta\rangle_{KP} = \frac{1}{\sqrt{N_{KP}(z;\beta)}}\exp\left[z\hat{\mathcal{A}}_+(\beta)\right]|0\rangle = \frac{1}{\sqrt{N_{KP}(z;\beta)}}\sum_{n=0}^{\infty}\frac{z^n}{n!}\left[\hat{\mathcal{A}}_+(\beta)\right]^n|0\rangle =$$

$$= \frac{1}{\sqrt{N_{KP}(z;\beta)}}\sum_{n=0}^{\infty}\frac{z^n}{n!}\sum_{k=0}^{n}\binom{n}{k}\left[\cosh\theta(\beta)\hat{\mathcal{A}}_+\right]^{n-k}\left[-\sinh\theta(\beta)\hat{\mathcal{A}}_-\right]^k|0\rangle \quad (4.50)$$

Taking into account result of the action of the annihilation operator on the vacuum state, the sum over the $k$ is different on zero only for $k = 0$. Then, we obtain

$$|z;\beta>_{KP} = \frac{1}{\sqrt{N_{KP}(z;\beta)}} \sum_{n=0}^{\infty} \frac{[z\cosh\theta(\beta)]^n}{n!} \left[\hat{\mathcal{A}}_+\right]^n |0> =$$

$$= \frac{1}{\sqrt{N_{KP}(z;\beta)}} \sum_{n=0}^{\infty} \frac{[z\cosh\theta(\beta)]^n}{n!} \sqrt{{}_p\rho_q(n)} |n> = \frac{1}{\sqrt{N_{KP}(z;\beta)}} \sum_{n=0}^{\infty} \frac{[z\cosh\theta(\beta)]^n}{\sqrt{{}_p\rho_q^{KP}(n)}} |n> \qquad (4.51)$$

Analyzing this expression, we find that it has the same mathematical form. The only difference is in the expression of the structure constants, which are now

$$_p\rho_q^{KP}(n) = \frac{(n!)^2}{{}_p\rho_q^{BG}(n)} = n! \frac{\prod_{i=1}^{p}(a_i)_n}{\prod_{j=1}^{q}(b_j)_n} \qquad (4.52)$$

From the normalization condition $<z;\beta|z;\beta>=1$ we obtain the expression of normalized function $N_{KP}(z;\beta)$

$$N_{KP}(z;\beta) = \sum_{n=0}^{\infty} \frac{\left[|z|^2 \cosh^2\theta(\beta)\right]^n}{{}_p\rho_q^{KP}(n)} = {}_qF_p\left(\boldsymbol{b}; \boldsymbol{a}; |z|^2 \cosh^2\theta(\beta)\right) \qquad (4.53)$$

Then, the normalized coherent state in "normal" Hilbert space becomes

$$|z;\beta>_{KP} = \frac{1}{\sqrt{{}_qF_p\left(\boldsymbol{b}; \boldsymbol{a}; |z|^2 \cosh^2\theta(\beta)\right)}} \sum_{n=0}^{\infty} \frac{[z\cosh\theta(\beta)]^n}{\sqrt{{}_p\rho_q^{KP}(n)}} |n> \qquad (4.54)$$

It can be seen that, compared to BG-CSs, in the case of KP-CSs there is an interchange between the indices $p$ and $q$, as well as between the sets of numbers $\boldsymbol{a}$ and $\boldsymbol{b}$.

It is observed that, at the limit

$$\lim_{\substack{T\to 0 \\ (\beta\to\infty)}} |z;\beta>_{KP} = \frac{1}{\sqrt{{}_qF_p\left(\boldsymbol{b}; \boldsymbol{a}; |z|^2\right)}} \sum_{n=0}^{\infty} \frac{z^n}{\sqrt{{}_p\rho_q^{KP}(n)}} |n> \equiv |z>_{KP} \qquad (4.55)$$

that is, the usual expression of KP-CSs, independent of temperature, is obtained.

The coherent states connected with the tilde space have the similar expression, so that the whole temperature dependent coherent states is

$$|z,\tilde{\sigma};\beta>_{KP} = |z;\beta>_{KP} \otimes |\tilde{\sigma};\beta>_{KP} =$$

$$= \left[ \frac{1}{\sqrt{{}_qF_p\left(\boldsymbol{b}; \boldsymbol{a}; |z|^2 \cosh^2\theta(\beta)\right)}} \sum_{n=0}^{\infty} \frac{[z\cosh\theta(\beta)]^n}{\sqrt{{}_p\rho_q^{KP}(n)}} |n> \right] \otimes \qquad (4.56)$$

$$\otimes \left[ \frac{1}{\sqrt{{}_qF_p\left(\boldsymbol{b}; \boldsymbol{a}; |\tilde{\sigma}|^2 \cosh^2\theta(\beta)\right)}} \sum_{n=0}^{\infty} \frac{[\tilde{\sigma}\cosh\theta(\beta)]^n}{\sqrt{{}_p\rho_q^{KP}(n)}} |\tilde{n}> \right]$$

The temperature dependent CSs, above constructed also fulfill all conditions imposed to an arbitrarily set of CSs, both of Barut-Girardello and Klauder-Perelomov. We will verify these properties for the Barut-Girardello type CSs, because for the Klauder-Perelomov type the proofs are similar. This is a consequence of the dualism that exists between the two types of CSs [32].

1.) *Continuity in the labels $z$ and $\tilde{\sigma}$.*

$$\lim_{\substack{z' \to z \\ \tilde{\sigma}' \to \tilde{\sigma}}} \| |z', \tilde{\sigma}'; \beta >_{BG} - |z, \tilde{\sigma}; \beta >_{BG} \| \to 0 \quad (4.57)$$

2.) *Normalization but non-orthogonality*. The overlap of CSs for physical Hilbert space is

$$_{BG}<z; \beta | z'; \beta>_{BG} = \frac{_pF_q(; z^*z' \cosh^2 \theta(\beta))}{\sqrt{_pF_q(; |z|^2 \cosh^2 \theta(\beta))} \sqrt{_pF_q(; |z'|^2 \cosh^2 \theta(\beta))}} =$$

$$= \begin{cases} 1 & \text{for } z = z' \quad \text{normalization} \\ \neq 0 & \text{for } z \neq z' \quad \text{non-orthogonality} \end{cases} \quad (4.58)$$

and similar for $|\tilde{\sigma}'; \beta>_{BG}$, so it results

$$_{BG}<z, \tilde{\sigma}; \beta | z', \tilde{\sigma}'; \beta>_{BG} = \begin{cases} 1 & \text{for } z = z' \text{ and } \tilde{\sigma} = \tilde{\sigma}' \quad \text{normalization} \\ \neq 0 & \text{for } z \neq z' \text{ and } \tilde{\sigma} \neq \tilde{\sigma}' \quad \text{non-orthogonality} \end{cases} \quad (4.59)$$

3.) *Over-completeness or unity operator decomposition*

$$\int d\mu_{BG}(z, \tilde{\sigma}; \beta) | z, \tilde{\sigma}; \beta><z, \tilde{\sigma}; \beta | = I \quad (4.60)$$

The whole integration measure is a product of two individual integration measures corresponding to the variables $z$ and $\tilde{\sigma}$. Moreover, the global integration measure is

$$d\mu_{BG}(z, \tilde{\sigma}; \beta) = d\mu_{BG}(z; \beta) \otimes d\mu_{BG}(\tilde{\sigma}; \beta) \quad (4.61)$$

so the over-completeness integral becomes

$$\int d\mu_{BG}(z; \beta) | z; \beta>_{BG} {}_{BG}<z; \beta | \otimes \int d\mu_{BG}(\tilde{\sigma}; \beta) | \tilde{\sigma}; \beta>_{BG} {}_{BG}<\tilde{\sigma}; \beta | = I \quad (4.62)$$

Substituting into the integral above, after some straightforward calculations it will be obtained that this integral is indeed equal to unity.

Therefore, the set of global CSs $|z, \tilde{\sigma}; \beta>$ satisfies all the requirements imposed on coherent states [19].

**4.2 Mixed states**

In usual quantum mechanics, the quantum states of a system can be either pure or mixed. But these two types of states do not benefit from the same type of description. While pure states are described by the state function (vector), mixed states are described by the density operator (matrix). The thermofield dynamics (TFD) formalism comes to correct this "injustice". In TFD, pure states and mixed states are treated equally, in the sense that each is represented by a state vector in the so-called "extended" Hilbert space, $H + \tilde{H}$. It includes, in addition to the physical Hilbert space $H$, also a virtual Hilbert space (or tilde) $\tilde{H}$.

However, in quantum mechanics, a *pure state* $|\Psi>$ in Hilbert space $H$ can be also described by a density operator. But, in this case, the density operator is just the projector of this state $\hat{\rho}_{|\Psi>} \equiv |\Psi><\Psi|$. Consequently, the expectation value of an observable $\hat{O}_H$ is calculated as the following trace

$$<\hat{O}_H>_{|\Psi>} = \text{Tr}(\hat{O}_H \hat{\rho}_{|\Psi>}) = \sum_n <n|\hat{O}_H|\Psi><\Psi|n> = <\Psi| \left( \sum_n |n><n| \right) \hat{O}_H |\Psi> = <\Psi|\hat{O}_H|\Psi>$$

(4.63)

So, the quantum statistics of a particle system (e.g. the bosons) in thermal equilibrium is described by the density operator. The canonical equilibrium states of a system at some non-zero temperature $T$ are the mixed states described by the thermodynamic density operator $\hat{\rho}(T) \equiv \hat{\rho}(\beta)$ defined as

$$\hat{\rho}(\beta) \equiv \frac{1}{Z(\beta)} \exp(-\beta \hat{\mathcal{H}}) = \frac{1}{Z(\beta)} \sum_{n=0}^{\infty} e^{-\beta E_n} |n><n| \quad , \quad Z(\beta) = \sum_{n=0}^{\infty} \exp(-\beta E_n) \qquad (4.64)$$

The thermal expectation value of an operator $\hat{O}_H$ is given by

$$<\hat{O}_H>_T = \text{Tr}[\hat{O}_H \hat{\rho}(\beta)] = \frac{1}{Z(\beta)} \sum_{n=0}^{\infty} e^{-\beta E_n} <n|\hat{O}_H|n> \qquad (4.65)$$

where the quantities $p_n(\beta) = [Z(\beta)]^{-1} e^{-\beta E_n}$ represent the probabilities of finding the physical system in the state $|n>$.

Consequently, the expectation values of an operator $\hat{O}_H$ acting on the physical Hilbert space $\mathsf{H}$, in a pure state $|n>$ is given by

$$<\hat{O}_H>_T = <n,\tilde{n}|\hat{O}_H|n,\tilde{n}> = <n|\hat{O}_H|n> \otimes <\tilde{n}|\tilde{n}> = <n|\hat{O}_H|n> \qquad (4.66)$$

If the system is at the zero temperature, $T \to 0$ (or $\beta \to \infty$), we have $p_n(\beta \to \infty) \to \delta_{n,0}$ and $\hat{\rho}(\beta \to \infty) \to |0><0|$, respectively $\hat{\rho}_{|0(\beta)>} = |0,\tilde{0}><0,\tilde{0}|$ for the whole Hilbert space.

Consequently, at the zero-temperature limit, the system will be in the pure fundamental vacuum state.

As a result of the repeated application of the creation operator on the vacuum state $|0>$, we will be able to write that the projector on the state $|n>$ is

$$|n><n| = \frac{1}{\#_p F_q(;\hat{\mathcal{A}}_+\hat{\mathcal{A}}_-)\#} \frac{\#(\hat{\mathcal{A}}_+\hat{\mathcal{A}}_-)^n \#}{_p\rho_q(n)} \qquad (4.67)$$

This leads to the expression for the density operator corresponding to a mixed state

$$\hat{\rho}_H(\beta) \equiv \frac{1}{Z_H(\beta)} \frac{1}{\#_p F_q(;\hat{\mathcal{A}}_+\hat{\mathcal{A}}_-)\#} \sum_{n=0}^{\infty} e^{-\beta E_n} \frac{\#(\hat{\mathcal{A}}_+\hat{\mathcal{A}}_-)^n \#}{_p\rho_q(n)} \qquad (4.68)$$

Similar expressions can be written for the density operator for the tilde field:

$$\hat{\tilde{\rho}}_{\tilde{H}}(\beta) \equiv \frac{1}{\tilde{Z}_{\tilde{H}}(\beta)} \frac{1}{\#_p F_q(;\hat{\tilde{\mathcal{A}}}_+\hat{\tilde{\mathcal{A}}}_-)\#} \sum_{n=0}^{\infty} e^{-\beta E_{\tilde{n}}} \frac{\#(\hat{\tilde{\mathcal{A}}}_+\hat{\tilde{\mathcal{A}}}_-)^n \#}{_p\rho_q(n)} \qquad (4.69)$$

Then, the whole density operator, corresponding to the both fields (original $\{|n>\}$ and tilde $\{|\tilde{n}>\}$) will be the tensor product of two individual density operators.

Using the properties of the tensor product, and assuming that the whole Hamiltonian, $\hat{\mathcal{H}}_{H\otimes\tilde{H}} = \hat{\mathcal{H}}_H \otimes \hat{\tilde{\mathcal{H}}}_{\tilde{H}}$ correspond to the combined Hilbert space $\mathsf{H}\otimes\tilde{\mathsf{H}}$, then the whole density operator can be written as

$$\hat{\rho}_{H\otimes\tilde{H}}(\beta)= \frac{1}{Z_{H\otimes\tilde{H}}(\beta)} \exp\left[-\beta\hat{\mathcal{H}}_{H\otimes\tilde{H}}\right] = \frac{1}{Z_{H\otimes\tilde{H}}(\beta)} \exp\left[-\beta\hat{\mathcal{H}}_{H\otimes\tilde{H}}\right]\sum_n |n,\tilde{n}><n,\tilde{n}|=$$

$$= \frac{1}{Z_H(\beta)} \exp\left[-\beta\hat{\mathcal{H}}\right]\sum_n |n><n| \otimes \frac{1}{Z_{\tilde{H}}(\beta)} \exp\left[-\beta\hat{\tilde{\mathcal{H}}}\right]\sum_n |\tilde{n}><\tilde{n}|= \quad (4.70)$$

$$= \left(\frac{1}{Z_H(\beta)}\sum_n e^{-\beta E_n}|n><n|\right)\otimes\left(\frac{1}{\tilde{Z}_{\tilde{H}}(\beta)}\sum_n e^{-\beta E_n}|\tilde{n}><\tilde{n}|\right) = \hat{\rho}_H(\beta)\otimes\hat{\tilde{\rho}}_{\tilde{H}}(\beta)$$

Previously we saw that the thermal vacuum state $|0(\beta)>$, which is a pure state of the whole or composite space is defined as

$$|0(\beta)>= \sum_n \sqrt{\frac{e^{-\beta E_n}}{Z_H(\beta)}}|n>\otimes|\tilde{n}> = \sum_n \sqrt{\frac{e^{-\beta E_n}}{Z_H(\beta)}}|n,\tilde{n}> \quad (4.71)$$

The bases $|n>$ and $|\tilde{n}>$ are called the Schmidt bases for Hilbert space $H$, and $\tilde{H}$ respectively, and consequently, the coefficients (called Schmidt coefficients) must be non negative and real numbers which satisfy the condition $\sum_n \left(\sqrt{\frac{e^{-\beta E_n}}{Z_H(\beta)}}\right)^2 = 1$ [33].

Additionally, we can transform the last row of the relation, thus

$$\frac{1}{Z_H(\beta)}\sum_n e^{-\beta E_n}|n,\tilde{n}><n,\tilde{n}| = \sum_n \sqrt{\frac{e^{-\beta E_n}}{Z_H(\beta)}}|n,\tilde{n}><n,\tilde{n}|\sqrt{\frac{e^{-\beta E_n}}{Z_H(\beta)}} \equiv \sum_n |n(\beta)><n(\beta)| \quad (4.72)$$

and similar for the fictitious space. The new used notation is

$$|n(\beta)> \equiv \sqrt{\frac{e^{-\beta E_n}}{Z_H(\beta)}}|n,\tilde{n}> \quad (4.73)$$

This can be interpreted as the temperature dependent Fock vectors. In this context we can define the following density operator for the pure $|n(\beta)>$ state:

$$\hat{\rho}_{|n(\beta)>} \equiv \sum_n |n(\beta)><n(\beta)| \quad (4.74)$$

Their trace is equal to unity $\text{Tr}\hat{\rho}_{|n(\beta)>} = 1$.

The whole energy of the system is $E_n^{total} = E_n + \tilde{E}_n$. But, since each real particle in the physical space $H$ has a fictitious counterpart (an image) in the fictitious space $\tilde{H}$, having the same energy, the result is that the total energy of the entire system is $2E_n$.

Then, the whole density operator can also be expressed in terms of the normal ordered product of temperature dependent ladder operators:

$$\hat{\rho}_{\mathsf{H}\otimes\tilde{\mathsf{H}}}(\beta) = \hat{\rho}_{\mathsf{H}}(\beta) \otimes \hat{\tilde{\rho}}_{\tilde{\mathsf{H}}}(\beta) =$$

$$= \left[ \frac{1}{Z_{\mathsf{H}}(\beta)} \frac{1}{\#_{p}F_{q}\left(;\hat{\mathcal{A}}_{+}\hat{\mathcal{A}}_{-}\right)\#} \sum_{n=0} e^{-\beta E_{n}} \frac{\#\left(\hat{\mathcal{A}}_{+}\hat{\mathcal{A}}_{-}\right)^{n}\#}{{}_{p}\rho_{q}(n)} \right] \quad (4.75)$$

$$\otimes \left[ \frac{1}{\tilde{Z}_{\tilde{\mathsf{H}}}(\beta)} \frac{1}{\#_{p}F_{q}\left(;\hat{\tilde{\mathcal{A}}}_{+}\hat{\tilde{\mathcal{A}}}_{-}\right)\#} \sum_{m=0} e^{-\beta E_{m}} \frac{\#\left(\hat{\tilde{\mathcal{A}}}_{+}\hat{\tilde{\mathcal{A}}}_{-}\right)^{m}\#}{{}_{p}\rho_{q}(m)} \right]$$

The trace of the density operator in the whole CSs representation is

$$\mathrm{Tr}\hat{\rho}_{\mathsf{H}\otimes\tilde{\mathsf{H}}}(\beta) = \int d\mu_{BG}(z,\tilde{\sigma};\beta)_{BG}\!<\!z,\tilde{\sigma};\beta|\hat{\rho}_{\mathsf{H}\otimes\tilde{\mathsf{H}}}(\beta)|z,\tilde{\sigma};\beta\!>_{BG} = \mathrm{Tr}\hat{\rho}_{\mathsf{H}}(\beta) \otimes \mathrm{Tr}_{\tilde{\mathsf{H}}}\hat{\rho}(\beta) \quad (4.76)$$

where the individual traces are, e.g.

$$\mathrm{Tr}\hat{\rho}_{\mathsf{H}}(\beta) = \int d\mu_{BG}(z;\beta)_{BG}\!<\!z;\beta|\hat{\rho}_{\mathsf{H}}(\beta)|z;\beta\!>_{BG} =$$

$$= \int d\mu_{BG}(z;\beta)\,\hat{\rho}_{\mathsf{H}}(\beta)\big|_{\#\hat{\mathcal{A}}_{+}\hat{\mathcal{A}}_{-}\#\to|z|^{2}} = \frac{1}{Z_{\mathsf{H}}(\beta)} \Gamma(a/b) \sum_{n} \frac{e^{-\beta E_{n}}}{{}_{p}\rho_{q}(n)} \times \quad (4.77)$$

$$\times \int_{0}^{\infty} d\big(|z|^{2}\cosh^{2}\theta(\beta)\big)\big(|z|^{2}\cosh^{2}\theta(\beta)\big)^{n} G_{p,\,q+1}^{q+1,0}\!\left(|z|^{2}\cosh^{2}\theta(\beta)\,\bigg|\,\begin{array}{c}/\,;\quad a\text{-}1\\0,\,b\text{-}1\,;\quad /\end{array}\right)$$

Since the result of the integration over the $|z|^2$ is $\left[\Gamma(a/b)\right]^{-1}{}_{p}\rho_{q}(n)$, finally we obtain that, as expected, $\mathrm{Tr}\hat{\rho}_{\mathsf{H}}(\beta)=1$, as well as $\mathrm{Tr}_{\tilde{\mathsf{H}}}\hat{\rho}(\beta)=1$. Then, $\mathrm{Tr}\hat{\rho}_{\mathsf{H}\otimes\tilde{\mathsf{H}}}(\beta)=1$.

Consequently, the expected value of an operator $\hat{O}_{\mathsf{H}}$ in a mixed state (the thermal average) in the tensor product space, in the whole CSs representation $|z,\tilde{\sigma};\beta>_{BG}$, is

$${}_{BG}\!<\!z,\tilde{\sigma};\beta|\hat{O}_{\mathsf{H}}|z,\tilde{\sigma};\beta\!>_{BG} = {}_{BG}\!<\!z;\beta|\hat{O}_{\mathsf{H}}|z;\beta\!>_{BG} \otimes \underbrace{{}_{BG}\!<\!\tilde{\sigma};\beta|\tilde{\sigma};\beta\!>_{BG}}_{=1} =$$

$$= \frac{\mathrm{Tr}\!\left(\hat{O}_{\mathsf{H}}\hat{\rho}_{\mathsf{H}\otimes\tilde{\mathsf{H}}}(\beta)\right)}{\mathrm{Tr}\hat{\rho}_{\mathsf{H}\otimes\tilde{\mathsf{H}}}(\beta)} = \mathrm{Tr}\!\left(\hat{O}_{\mathsf{H}}\hat{\rho}_{\mathsf{H}\otimes\tilde{\mathsf{H}}}(\beta)\right) = \mathrm{Tr}\hat{O}_{\mathsf{H}} \otimes \underbrace{\mathrm{Tr}\!\left(\hat{\tilde{\rho}}_{\tilde{\mathsf{H}}}(\beta)\right)}_{=1} = \mathrm{Tr}\hat{O}_{\mathsf{H}} = \sum_{n}\!<\!n|\hat{O}_{\mathsf{H}}|n\!>=<\hat{O}_{\mathsf{H}}> \quad (4.78)$$

It is known that in quantum optics *the Q-distribution function* (also known as Husimi's Q-function) is defined as being equal to the diagonal elements of the density operator in the representation of coherent states [34].

For the case of whole coherent states we have

$$Q_{\mathsf{H}+\tilde{\mathsf{H}}}\!\left(|z|^{2},|\tilde{\sigma}|^{2};\beta\right) \equiv <\!z,\tilde{\sigma};\beta|\hat{\rho}_{\mathsf{H}+\tilde{\mathsf{H}}}(\beta)|z,\tilde{\sigma};\beta\!>_{BG} =$$

$$= {}_{BG}\!<\!z;\beta|_{BG}\hat{\rho}_{\mathsf{H}}(\beta)|z;\beta\!>_{BG} \otimes <\!\tilde{\sigma};\beta|\hat{\rho}_{\tilde{\mathsf{H}}}(\beta)|\tilde{\sigma};\beta\!>_{BG} =$$

$$= \frac{1}{Z_{\mathsf{H}}(\beta)} \frac{1}{{}_{p}F_{q}\left(;|z|^{2}\cosh^{2}\theta(\beta)\right)} \sum_{n} e^{-\beta E_{n}} \frac{\left[|z|^{2}\cosh^{2}\theta(\beta)\right]^{n}}{{}_{p}\rho_{q}(n)} \otimes \quad (4.79)$$

$$\otimes \frac{1}{Z_{\tilde{\mathsf{H}}}(\beta)} \frac{1}{{}_{p}F_{q}\left(;|\tilde{\sigma}|^{2}\cosh^{2}\theta(\beta)\right)} \sum_{n} e^{-\beta E_{n}} \frac{\left[|\tilde{\sigma}|^{2}\cosh^{2}\theta(\beta)\right]^{n}}{{}_{p}\rho_{q}(n)} =$$

$$= Q_{\mathsf{H}}\!\left(|z|^{2};\beta\right) \otimes Q_{\tilde{\mathsf{H}}}\!\left(|\tilde{\sigma}|^{2};\beta\right)$$

The *diagonal representation* of the integer density operator in the CSs representation can be written as

$$\hat{\rho}_{H\otimes\tilde{H}}(\beta) = \int d\mu(z,\tilde{\sigma};\beta)|z,\tilde{\sigma};\beta>_{BG} P_{H\otimes\tilde{H}}(|z|^2,|\tilde{\sigma}|^2;\beta)_{BG}<z,\tilde{\sigma};\beta| \quad (4.80)$$

Due to the symmetry of two variables, $z$ and $\tilde{\sigma}$, the whole $P$-quasi distribution function $P_{H\otimes\tilde{H}}(|z|^2,|\tilde{\sigma}|^2;\beta)$ can be split into two parts:

$$P_{H\otimes\tilde{H}}(|z|^2,|\tilde{\sigma}|^2;\beta) = P_H(|z|^2;\beta) P_{\tilde{H}}(|\tilde{\sigma}|^2;\beta) \quad (4.81)$$

so that we obtain

$$\hat{\rho}_{H\otimes\tilde{H}}(\beta) = \hat{\rho}_H(\beta)\otimes\hat{\rho}_{\tilde{H}}(\beta) =$$
$$= \int d\mu(z;\beta)|z;\beta>_{BG} P_H(|z|^2;\beta)_{BG}<z;\beta| \otimes \int d\mu(\tilde{\sigma};\beta)|\tilde{\sigma};\beta>_{BG} P_{\tilde{H}}(|\tilde{\sigma}|^2;\beta)_{BG}<\tilde{\sigma};\beta| \quad (4.82)$$

On the other hand, the whole density operator can be expressed as

$$\hat{\rho}_{H\otimes\tilde{H}}(\beta) = \hat{\rho}_H(\beta) \otimes \hat{\rho}_{\tilde{H}}(\beta) =$$
$$= \left(\frac{1}{Z_H(\beta)}\sum_{n=0}^{\infty} e^{-\beta E_n}|n><n|\right) \otimes \left(\frac{1}{\tilde{Z}_{\tilde{H}}(\beta)}\sum_{\tilde{n}=0}^{\infty} e^{-\beta E_{\tilde{n}}}|\tilde{n}><\tilde{n}|\right) \quad (4.83)$$

Equating the two expressions, we will have, for example for the variable $z$

$$\int d\mu(z;\beta) P_H(|z|^2;\beta)|z;\beta>_{BG}\,_{BG}<z;\beta| = \frac{1}{Z_H(\beta)}\sum_{n=0}^{\infty} e^{-\beta E_n}|n><n| \quad (4.84)$$

If we make explicit the expressions of coherent states, we will have

$$\Gamma(a/b)\sum_{n,m}\frac{[\cosh\theta(\beta)]^n}{\sqrt{_p\rho_q^{BG}(n)}}|n><m|\frac{[\cosh\theta(\beta)]^m}{\sqrt{_p\rho_q^{BG}(m)}}\times$$
$$\times\int_0^{\infty} d(|z|^2\cosh^2\theta(\beta)) P_H(|z|^2;\beta) G_{p,q+1}^{q+1,0}\left(|z|^2\cosh^2\theta(\beta)\,\bigg|\begin{array}{c}/\,;\quad a-1\\0\,,\,b-1\,;\quad/\end{array}\right)\times \quad (4.85)$$
$$\times\int_0^{2\pi}\frac{d\varphi}{2\pi} z^n(z^*)^m = \frac{1}{Z_H(\beta)}\sum_{n=0}^{\infty} e^{-\beta E_n}|n><n|$$

The result of the angular integral being $(|z|^2)^n \delta_{nm}$, we now have

$$\Gamma(a/b)\sum_n \frac{1}{_p\rho_q^{BG}(n)}|n><n|\times$$
$$\times\int_0^{\infty} d(|z|^2\cosh^2\theta(\beta)) P_H(|z|^2;\beta) G_{p,q+1}^{q+1,0}\left(|z|^2\cosh^2\theta(\beta)\,\bigg|\begin{array}{c}/\,;\quad a-1\\0\,,\,b-1\,;\quad/\end{array}\right)[|z|^2\cosh^2\theta(\beta)]^n =$$
$$= \sum_{n=0}^{\infty}\frac{e^{-\beta E_n}}{Z_H(\beta)}|n><n|$$

(4.86)

It follows that the integral must be equal to

$$\int_0^\infty d(|z|^2 \cosh^2\theta(\beta)) P_H(|z|^2;\beta) G_{p,q+1}^{q+1,0}\left(|z|^2 \cosh^2\theta(\beta) \left|\begin{array}{c} /\,; \quad a-1 \\ 0,\, b-1\,; \quad / \end{array}\right.\right) [|z|^2 \cosh^2\theta(\beta)]^n =$$

$$= \frac{1}{\Gamma(a/b)} \frac{1}{Z_H(\beta)} e^{-\beta E_n} \,_p\rho_q^{BG}(n) \tag{4.87}$$

Continuing the problem, i.e. finding the expression, can only be done if we know the concrete expression of the energy eigenvalues, as functions of the principal quantum number $n$.

For example, for systems with a linear energy spectrum, $E_n = \hbar\omega n + E_0$, for which the partition function is $Z_H(\beta) = e^{-\beta E_0}(1 - e^{-\beta\hbar\omega})^{-1}$. Then, the solution process proceeds as follows: Firstly, we denote the desired function with $S_H(|z|^2;\beta)$ and change the exponent. $n = s-1$:

$$S_H(|z|^2;\beta) \equiv G_{p,q+1}^{q+1,0}\left(|z|^2 \cosh^2\theta(\beta)\left|\begin{array}{c} /\,; \quad a-1 \\ 0,\, b-1\,; \quad / \end{array}\right.\right) P_H(|z|^2;\beta) \tag{4.88}$$

So, we have

$$\int_0^\infty d(|z|^2 \cosh^2\theta(\beta)) S_H(|z|^2;\beta)(|z|^2 \cosh^2\theta(\beta))^{s-1} =$$

$$= \frac{1}{\Gamma(a/b)} \frac{e^{-\beta E_n}}{Z_H(\beta)} (e^{-\beta\hbar\omega})^n \,_p\rho_q^{BG}(n) = (1-e^{-\beta\hbar\omega})e^{\beta\hbar\omega} \frac{1}{(e^{\beta\hbar\omega})^s} \Gamma(s) \frac{\prod_{j=1}^q \Gamma(b_j - 1 + s)}{\prod_{i=1}^p \Gamma(a_i - 1 + s)} \tag{4.89}$$

This is again a Stieltjes moment problem, the solution of which is

$$S_H(|z|^2;\beta) = (e^{\beta\hbar\omega} - 1) G_{p,q+1}^{q+1,0}\left(|z|^2 e^{\beta\hbar\omega} \cosh^2\theta(\beta)\left|\begin{array}{c} /\,; \quad a-1 \\ 0,\, b-1\,; \quad / \end{array}\right.\right) \tag{4.90}$$

Finally, the *P*-quasi-distribution function for the variable z is

$$P_H(|z|^2;\beta) =$$

$$= (e^{\beta\hbar\omega} - 1) \frac{G_{p,q+1}^{q+1,0}\left(|z|^2 e^{\beta\hbar\omega} \cosh^2\theta(\beta)\left|\begin{array}{c} /\,; \quad a-1 \\ 0,\, b-1\,; \quad / \end{array}\right.\right)}{G_{p,q+1}^{q+1,0}\left(|z|^2 \cosh^2\theta(\beta)\left|\begin{array}{c} /\,; \quad a-1 \\ 0,\, b-1\,; \quad / \end{array}\right.\right)} \tag{4.91}$$

In the limit, for $\theta(\beta) \to 0$, and using distribution formula for the expected number of particles in an energy state $|n>$ for Bose–Einstein statistics

$$\bar{n} = \frac{1}{e^{\beta\hbar\omega} - 1}, \quad \cosh^2\theta(\beta) = \bar{n} + 1 \tag{4.92}$$

then the above expression is identical to Eq. (4.19) of [27]. Similar expression is obtained for the variable $\tilde{\sigma}$, and the whole *P*-quasi-distribution function for the extended system with linear energy spectra is, finally

$$P_{H+\tilde{H}}(|z|^2,|\tilde{\sigma}|^2;\beta) = P_H(|z|^2;\beta) P_{\tilde{H}}(|\tilde{\sigma}|^2;\beta) \tag{4.93}$$

The expected value of an operator $\hat{O}_H$ acting on the physical Hilbert space H can be expressed using the diagonal representation of the density operator

$$<\hat{O}_H>_H = \text{Tr}_{H+\tilde{H}}\left[\hat{O}_H \hat{\rho}_{H+\tilde{H}}(\beta)\right] = \text{Tr}_H\left[\hat{O}_H \hat{\rho}_H(\beta)\right] \otimes \underbrace{\text{Tr}_{\tilde{H}}\left[\hat{\rho}_{\tilde{H}}(\beta)\right]}_{=1} = \text{Tr}_H\left[\hat{O}_H \hat{\rho}_H(\beta)\right] =$$

$$= \int d\mu(z;\beta) <z;\beta|\hat{O}_H \hat{\rho}_H(\beta)|z;\beta> =$$

$$= \int d\mu(z;\beta) <z;\beta|\hat{O}_H \left[\int d\mu(z';\beta) P_H(|z'|^2;\beta)|z';\beta><z';\beta|\right]|z;\beta> =$$

$$= \int d\mu(z';\beta) P_H(|z'|^2;\beta) <z';\beta|\hat{O}_H|z';\beta> \tag{4.94}$$

The expected value of the operator $\hat{O}_H$ in the coherent state $|z';\beta>$ can be written as

$$<z';\beta|\hat{O}_H|z';\beta> = \sum_{n,m} \frac{1}{\sqrt{{}_p\rho_q^{BG}(n)}} \frac{1}{\sqrt{{}_p\rho_q^{BG}(m)}} \frac{[z\cosh\theta(\beta)]^n [z^*\cosh\theta(\beta)]^m}{{}_pF_q(;|z'|^2\cosh^2\theta(\beta))} <m|\hat{O}_H|n> \tag{4.95}$$

Replacing in the previous relation it will be obtained that $m=n$, we will be able to use the result of Eq. (PG int) and we will obtain the expression of thermal average, as we expected

$$<\hat{O}_H>_H = \frac{1}{Z_H(\beta)} \sum_{n=0}^{\infty} e^{-\beta E_n} <n|\hat{O}_H|n> = <\hat{O}_H>_T \tag{4.96}$$

Let us now show that the expected value of the $\hat{O}_H$ operator in the pure thermal vacuum state is identical (equal) to the thermal average value of the $\hat{O}_H$ operator in the physical Hilbert space H.

Knowing that the calculation of the trace does not depend on the base used, for verification we will perform the calculations in two different bases:

1. Using the Fock vector basis for the entire system, $|n,\tilde{n}> = |n>\otimes|\tilde{n}>$

$$<0(\beta)|\hat{O}_H|0(\beta)> = \sum_n <n,\tilde{n}|\hat{\rho}_{H+\tilde{H}}(\beta)\hat{O}_H|n,\tilde{n}> = \sum_n <n|\otimes<\tilde{n}|\hat{\rho}_{H+\tilde{H}}(\beta)\hat{O}_H|n>\otimes|\tilde{n}> =$$

$$= \sum_n <n|\otimes<\tilde{n}|\left[\hat{\rho}_H(\beta)\otimes\hat{\tilde{\rho}}_{\tilde{H}}(\beta)\right]\hat{O}_H|n>\otimes|\tilde{n}> = \sum_n <n|\hat{\rho}_H(\beta)\hat{O}_H|n>\otimes<\tilde{n}|\hat{\tilde{\rho}}_{\tilde{H}}(\beta)|\tilde{n}> \tag{4.97}$$

$$<0(\beta)|\hat{O}_H|0(\beta)> = \sum_n <n|\hat{\rho}_H(\beta)\hat{O}_H|n>\otimes I = \sum_n <n|\hat{\rho}_H(\beta)\hat{O}_H|n> =$$

$$= \sum_n \frac{1}{Z_{\tilde{H}}(\beta)} \sum_m e^{-\beta E_m} <n|m><m|\hat{O}_H|n> = \frac{1}{Z_{\tilde{H}}(\beta)} \sum_n e^{-\beta E_n} <n|\hat{O}_H|n> = <\hat{O}_H>_T \tag{4.98}$$

So, finally

$$<0(\beta)|\hat{O}_H|0(\beta)> = <\hat{O}_H>_T \tag{4.99}$$

2. Using the coherent states vector basis for the entire system (physical and tilda), $|z,\tilde{\sigma};\beta> = |z;\beta>\otimes|\tilde{\sigma};\beta>$.

$$<0(\beta)|\hat{O}_H|0(\beta)> = \int d\mu(z,\tilde{\sigma};\beta)<z,\tilde{\sigma};\beta|\hat{\rho}_{H+\tilde{H}}(\beta)\hat{O}_H|z,\tilde{\sigma};\beta> =$$
$$= \int d\mu(z;\beta)<z;\beta|\hat{\rho}_H(\beta)\hat{O}_H|z;\beta> \otimes \int d\mu(\tilde{\sigma};\beta)<\tilde{\sigma};\beta|\hat{\tilde{\rho}}_{\tilde{H}}(\beta)|z,\tilde{\sigma};\beta> =$$
$$= \int d\mu(z;\beta)<z;\beta|\hat{\rho}_H(\beta)\hat{O}_H|z;\beta> =$$
$$= \int d\mu(z';\beta)P_H(|z'|^2;\beta)<z';\beta|\hat{O}_H|z';\beta> = <\hat{O}_H> \quad (4.100)$$

where we used the result of Eq. (4.94).

## 5. Concluding remarks

As we showed in the introduction, the thermofield dynamics (TFD) has proven to be an important approach, due to its multiple applications in different branches of physics, where temperature or time-dependent phenomena occur. On the other hand, the formalism of coherent states (CSs), initially imagined only as the prerogative of quantum optics, over time has found its place in other fields, to highlight just quantum computing. Therefore, the synergistic link between TFD and CSs is a topic that is as interesting as it is exciting.

In the paper we tried to show that the TFD rules, formulated for linear (canonical) operators, can be extended to nonlinear or, as they are called - deformed operators, associated with systems of deformed bosons. Although in the paper we only dealt with boson systems, the discussion can also be extended to fermion systems. We built sets of temperature dependent coherent states, both of the Barut-Girardello and of the Klauder-Perelomov type, these two types being dual to each other. In calculations involving ladder operators, both independent and temperature-dependent, we used the Diagonal Operator Ordering Technique (DOOT), which is a generalization of the Integration Within an Ordered Product technique (IWOP) applicable to canonical operators associated with the one-dimensional harmonic oscillator.

Before addressing the issue of coherent states, we highlighted the connection between the expected value of the normal ordered product of operators in the thermal vacuum state and the value of the internal energy of the bosonic "gas"

The construction of the thermal coherent states (TCSs) was done through a natural extension, taking into account that the algebras of the thermal operators and those at zero-temperature are identical. This was done within the thermo field dynamics (TFD) formalism, which essentially represents an extension of the field theory from zero-temperatures to finite temperatures. Because thermal coherent states (TCSs), in addition to physical degrees of freedom, also contain thermal degrees of freedom, it was necessary to expand the physical Hilbert space $H$, adding a virtual Hilbert space (or tilde) $\tilde{H}$. Consequently, we paid increased attention to the construction and properties of the corresponding density operators, responsible for both pure and mixed quantum states.

The presented TFD formalism, formulated for temperature-dependent quantities, can be extended also to the time-dependent phenomena, through the formal substitution $\beta \rightarrow i(\hbar)^{-1}t$, so that $\exp(-\beta\hat{\mathcal{H}}) \rightarrow \exp\left(-\frac{i}{\hbar}\hat{\mathcal{H}}t\right)$ [35].

Finally, for informational purposes, let's mention one of the newest applications of TFD, namely how this approach can be connected to quantum computing. If we consider the particular case of a two-level system, the expression for the thermal vacuum will be

$$|0(\beta)>=\sum_{n=0}^{1}\sqrt{\frac{e^{-\beta E_n}}{Z(\beta)}}|n,\tilde{n}>=\sqrt{\frac{e^{-\beta E_0}}{Z(\beta)}}|0,\tilde{0}>+\sqrt{\frac{e^{-\beta E_1}}{Z(\beta)}}|1,\tilde{1}> \qquad (5.1)$$

and the partition function is $Z(\beta)=e^{-\beta E_0}+e^{-\beta E_1}$, so that finally we have

$$|0(\beta)>=\frac{1}{\sqrt{1+e^{-\beta(E_1-E_0)}}}|0,\tilde{0}>+\frac{1}{\sqrt{1+e^{+\beta(E_1-E_0)}}}|1,\tilde{1}> \qquad (5.2)$$

Since the sum of the squares of the coefficients is equal to 1, if we identify $|0,\tilde{0}>\equiv|0>$ and $|1,\tilde{1}>\equiv|1>$, the expression for the thermal vacuum $|0(\beta)>$ becomes identical to the new defined expression for the thermal vacuum qubit or thermal vacuum:

$$|\Psi(\beta)>=c_0|0>+c_1|1> \quad , \quad c_0^2+c_1^2=1 \qquad (5.3)$$

This clearly shows the connection between thermofield dynamics and quantum computing. For details we recommend the reader to the recent papers [36], [37] (and references therein).

We hope that the results presented here will constitute a step forward in diversifying applications involving temperature- or time-dependent phenomena and systems, both for TFD and for CSs.